\documentclass[twocolumn,floatfix,trackchanges]{aastex631}
\usepackage{hyperref}

\usepackage{amsmath}

\usepackage{float}
\usepackage{soul}
\usepackage{needspace}

\newcommand\N[1]{{\color{blue} \bf #1}}

\newcommand{\eq}[1]{$\mathrm{#1}$}

\shortauthors{Hillenkamp et al.}
\newcommand{\Ms}{M$_{\odot}$}

\begin{document}

% \setcapwidth{0.95\textwidth}
% \setLTcapwidth{0.95\textwidth}

\title{Fading Echoes of Interaction: Probing Centuries of Preexplosion Mass-Loss in Four Type IIn Supernovae}

\author[0009-0000-8192-1204]{Elizabeth Hillenkamp}
\affiliation{Department of Astronomy \& Astrophysics, University of California, San Diego, 9500 Gilman Dr., La Jolla, CA 92093, USA}
\affiliation{National Radio Astronomy Observatory, 520 Edgemont Rd, Charlottesville, VA 22903, USA}
\email{ehillenkamp@ucsd.edu}

\author[0009-0004-7268-7283]{Raphael Baer-Way}
\affiliation{National Radio Astronomy Observatory, 520 Edgemont Rd, Charlottesville, VA 22903, USA}

\affiliation{Department of Astronomy, University of Virginia, Charlottesville, VA 22904-4325, USA}

\author[0000-0002-0844-6563]{Poonam Chandra}
\affiliation{National Radio Astronomy Observatory, 520 Edgemont Rd, Charlottesville, VA 22903, USA}

\author[0000-0002-9820-679X]{Arkaprabha Sarangi}\affiliation{Indian Institute of Astrophysics, 
100 Feet Rd, Koramangala, Bengaluru, Karnataka 560034, India}

%\email[show]{arkaprabha.sarangi@iiap.res.in}

\author[0000-0002-9117-7244]{Roger Chevalier}\affiliation{Department of Astronomy, University of Virginia, Charlottesville, VA 22904-4325, USA}

\author[0000-0002-8070-5400]{Nayana A.J.}
\affiliation{Department of Astronomy, University of California, Berkeley, CA 94720-3411, USA}
\affiliation{Berkeley Center for Multi-messenger Research on Astrophysical Transients and Outreach (Multi-RAPTOR), University of California, Berkeley, CA 94720-3411, USA}

\author[0009-0000-5307-8897]{Annika Deutsch}\affiliation{Department of Astronomy, University of Virginia, Charlottesville, VA 22904-4325, USA}

\author[0000-0003-2611-7269]{Keiichi Maeda}\affiliation{Department of Astronomy, Kyoto University, Sakyo-ku, Kyoto, 606-8502, Japan}

\author[0000-0001-5510-2424]{Nathan Smith}\affil{Steward Observatory, University of Arizona, 933 N. Cherry Avenue, Tucson, AZ 85721, USA}



\begin{abstract}

Supernovae characterized by enduring narrow optical hydrogen emission lines (SNe IIn) are believed to result primarily from the core-collapse of massive stars undergoing sustained interaction with a dense circumstellar medium (CSM). While the properties of SN IIn progenitors have relatively few direct constraints, the ongoing ejecta-CSM interaction provides unique information about late-stage stellar mass-loss preceding core-collapse. We present late-time X-ray and radio observations of four $\geq$3000 day-old SNe IIn: SN 2013L, SN 2014ab, SN 2015da, and KISS15s. The radio and X-ray emission from KISS15s indicate a mass-loss rate of \eq{\dot M\sim4\times 10^{-3}~\rm{M_{\odot}\,yr^{-1}}} at $\sim$450 years pre-supernova --- 2 orders of magnitude below earlier optical estimates (which probed the mass loss immediately preceding the supernova). We find hints of a spectral inversion in the radio SED of KISS15s; a possible signature of a secondary shock due to a binary system or the emergence of a pulsar wind. For SN 2013L, we obtain a mass-loss rate of \eq{\dot M\sim2 \times 10^{-3}~\rm{M_{\odot}\,yr^{-1}}} at $\sim$400 years pre-explosion based on the X-ray detection. For SN 2014ab and SN 2015da, we find a upper limits on the mass-loss rates of \eq{\dot M<2\times10^{-3}~M_{\sun}\,yr^{-1}} explosion at $\sim$300 and 250 years pre-explosion, respectively. All four objects display mass-loss rates lower than estimates from earlier optical analyses by at least 1-2 orders of magnitude, necessitating a rapidly evolving progenitor process over the last centuries pre-explosion. Our analysis reveals how X-ray and radio observations can elucidate progenitor evolution when these objects have faded at optical wavelengths.

\end{abstract}

\keywords{Stellar mass loss (1613) --- Core-collapse supernovae (304)  ---  Circumstellar matter (241)  --- X-ray transient sources (1852) --- Extragalactic radio sources (508)} 

\section{Introduction} \label{sec:intro}

\begin{figure}
    \centering
    \includegraphics[width=.45\textwidth]{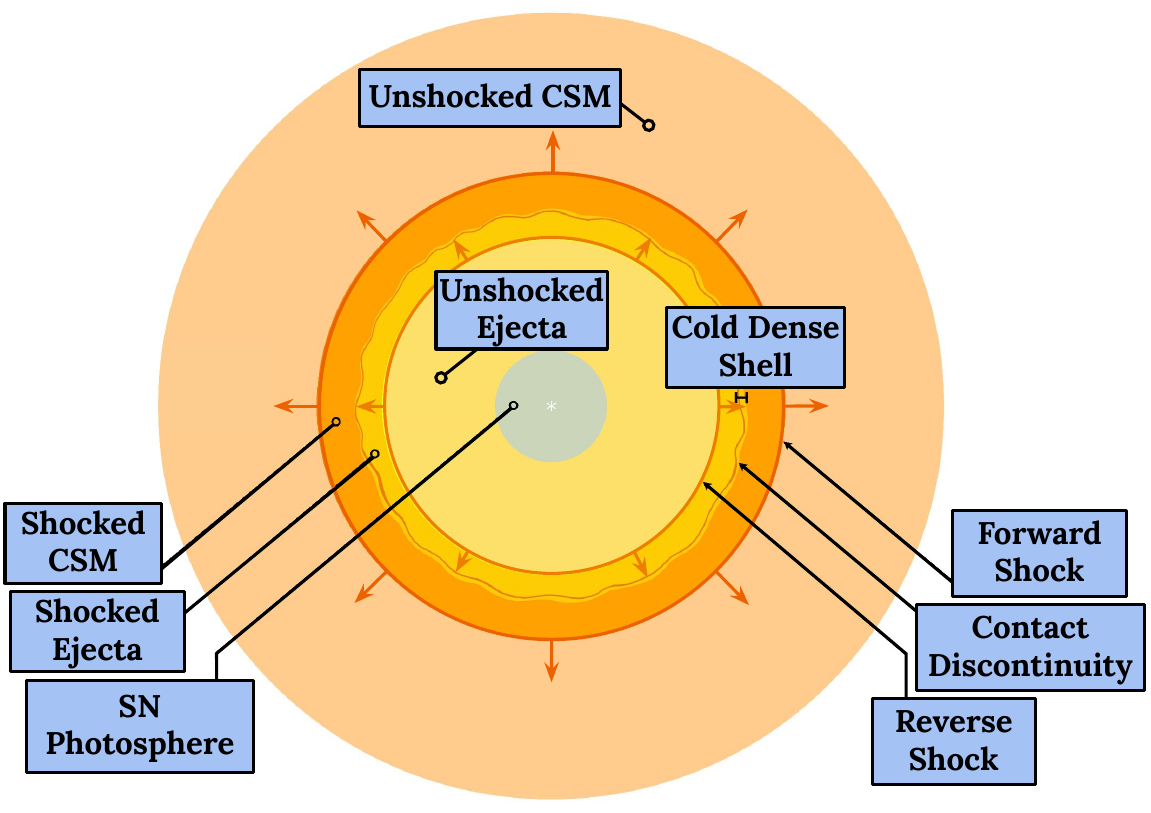} 
    \caption{A (simplified and not to scale) model of circumstellar interaction in a Type IIn supernova. Narrow optical emission lines originate in the unshocked photoionized CSM, while broad optical emission lines are produced by supernova ejecta. The ejecta interaction with the CSM creates a forward and reverse shock. The hot forward shock can accelerate the CSM to very high speeds, producing X-ray and radio synchrotron emission, which may be reprocessed into UV and optical. At late times, X-ray and UV emission from the cooler reverse shock dominates. Based on \cite{Chandra2018}.} 
    \label{fig:SN_diagram}
\end{figure}

Type IIn supernovae (SNe IIn), first identified by \citet{Schlegel1990}, are a class of supernovae distinguished by narrow hydrogen emission lines in their early optical spectra and slowly evolving blue continua. SNe IIn are relatively rare; making up from 5\% \citep{Cold2023} to 9\% \citep{smith11frac,Li2011} of core-collapse supernovae. The characteristic narrow emission lines are believed to result from a relatively slow-moving photoionized circumstellar medium (CSM) formed from material that was blown off of the progenitor star during the final stages of its life \citep{Smith2017}. When the ejecta encounters the CSM, forward and reverse shocks form due to the difference in ejecta ($\sim$ \eq{10,000~km\,s^{-1}}) and CSM ($\sim$ \eq{100~km\,s^{-1}}) speeds. The forward shock propagates through the CSM, while the slower reverse shock passes through the ejecta, moving inwards with respect to the forward shock but outwards in the observer frame (see Figure \ref{fig:SN_diagram} for a visual representation). In some cases, a cold dense shell (CDS) may form at the site of the contact discontinuity in between the shocks, which may provide an environment for dust formation, as proposed by \citet{Pozzo2004} (seen in SN 1998S and other SNe IIn as a strong late-time infrared excess). The ejecta-CSM interaction itself produces radiation that helps constrain the pre-explosion mass-loss rate of the progenitor star, thereby probing its late-stage evolution. 

% Possible progenitors & caveats (1 para) 
High mass-loss rates (\eq{\gtrsim 10^{-2}~{M_{\odot}\,yr^{-1}}}) are a ubiquitous feature of the progenitors of SNe IIn \citep{Smith2014}. The mass-loss rates observed in SNe IIn can only be explained by progenitor stars with sustained or episodic periods of very high mass loss prior to explosion \citep{Smith_2007}. Most individual massive stars such as super-asymptotic giant branch stars (\eq{8 - 10\,M_{\sun}}), red supergiants (RSGs; \eq{\sim17-40\,M_{\sun}}), and yellow hypergiants are only expected to produce enhanced winds up to \eq{10^{-4}\sim10^{-3}~M_{\sun}\,yr^{-1}}, which is not as high as the majority of measured SN IIn mass-loss rates \citep{Smith2014}. Luminous blue variable (LBV; \eq{>25M_{\sun}}) stars \citep{smith26} can produce the requisite rates with episodic mass loss of \eq{\sim0.1~M_{\sun}\,yr^{-1}} \citep{Smith2014, Chandra2018, Chandra_2025}. The supernova channel for LBVs is surprising or even problematic in the standard single-star view of massive star evolution, in which LBVs have been believed to be a transitional phase between O-type and Wolf-Rayet (W-R) stars \citep[see e.g., contributions in][]{nota97}. However, there has been compelling evidence for certain objects exploding in the LBV phase itself \citep[e.g.][]{Smith_2007,Galyam_2007,trundle08,smith10ip,Mauerhan_2013,Smith2017,Nagao_2025}.

Beyond the unifying factors of prolonged emission and persisting narrow hydrogen lines, SNe IIn display a wide diversity in their spectral properties and lightcurve evolution \citep{Ransome_2025}. As observational campaigns continue to populate the parameter space, evidence accrues for possible subclasses of SNe IIn, perhaps even resulting from distinct progenitors. Previous works have identified anywhere from 3 \citep{Taddia2013} to 7 or more \citep{Smith2017} informal subtypes of SNe IIn based on the development of spectral characteristics. A few of the most common subclassifications are discussed below.

The first proposed subtype (1988Z-like) is perhaps the most classical picture of SNe IIn, defined by enduring emission that can persist for decades due to interaction with a CSM formed from a mass-loss rate $> 10^{-3}\,\rm{M_{\odot}~yr^{-1}}$. The lightcurves are powered by ongoing CSM interaction, indicative of mass-loss events that lasted hundreds or thousands of years before core collapse. The mass-loss rates typical of these SNe may be explained by enhanced winds from RSGs \citep{Smith2017}.

Another group of note are the superluminous Type IIn supernovae (SLSNe IIn), which have been observed with evidence for both extended shells as in SN 2010jl and compact shells as in the case of SN 2006gy \citep{Smith_2007,Zhang_10jl}. SLSNe IIn display notable lightcurve diversity, but in general, the mass-loss rates required for these supernovae are too high to result from red giants and are commonly linked to LBV eruptions or other episodic mass-loss events \citep{Chandra2015, Smith_2017}. As mentioned above, LBVs pose an intriguing progenitor candidate --- if they are indeed transitional phase stars, they are not expected to go supernova; however, \citet{Smith2015} found statistical variation in location and isolation of LBVs compared with OB and W-R stars, suggesting that LBVs may instead be members of massive binaries which grow through merger events or mass transfer before being kicked from the system by their companion's supernova.

Binary systems themselves are increasingly invoked as progenitor scenarios for many SNe IIn, especially those with clear evidence for asymmetric CSM \citep[see e.g.][]{bilinski24,Smith2024,BaerWay2025}. This is unsurprising given recent results suggesting the majority of massive stars end up in binary systems \citep{Sana_2012}. 

Other commonly discussed subclasses include transitional SNe IIn such as SN 1998S \citep{Shivvers_2015}, which display fleeting narrow hydrogen lines; delayed onset SNe IIn such as SN 2009ip \citep{Mauerhan_2013}; and Type IIn-P SNe such as 1994W amd 2011ht \citep{Chugai_2004,mauerhan13}, which display a clear plateau along with signatures of CSM interaction \citep{smith13}. Thus, SNe IIn as a whole may only be unified by their notably dense CSM, rather than a common class of progenitors.

%Emission from SNe IIn 
The early optical spectroscopic studies of these objects provide insight regarding their wind, ejecta, and shock velocities. The wind (or CSM) speed can only be constrained through the narrow components of the hydrogen emission lines ($\mathrm{v_{FWHM}} \sim$\eq{100~km\,s^{-1}}). These measurements are especially robust when P Cygni profiles can be resolved. Intermediate-width emission line components ($\mathrm{v_{FWHM}} \sim$\eq{2000~km\,s^{-1}}) have two different origins: they are generally believed to be formed by electron scattering wings from the dense inner CSM at early times, while at later times (post-peak), they arise in the CSM-ejecta interaction region between the forward and reverse shocks, and can provide information about the forward and reverse shock velocities. Broad line components ($\mathrm{v_{FWHM}}$ \eq{\gtrsim10,000~\mathrm{km\,s}^{-1}}) are sometimes cited as a probe of the unshocked ejecta velocity, seen due to asymmetries \citep{BaerWay2025}, but in other cases the emission likely originates in the pre- or post-shock gas and has been subsequently broadened by multiple electron-scattering events in the CSM \citep[e.g.][]{Chugai2001, Huang2018}. 

The progenitor star mass-loss rate can be estimated from the SN bolometric luminosity combined with constraints on the CSM and shock velocities  \citep{Smith2017}. As shown by \citet{Chugai1991}, the \eq{H\alpha} luminosity can serve as a reliable proxy of the bolometric luminosity when the SN emission is dominated by CSM interaction, and can thus also be used to estimate the progenitor mass-loss rate. However, H$\alpha$ mass-loss rate calculations rely on the assumption of a smooth, steady wind and may overestimate the mass-loss rate due to factors such as clumpiness \citep[e.g.][]{Owocki1988, Feldmeier1995, Fullerton2006}, whereas mass-loss rates estimated from the continuum luminosity during the main light-curve peak are usually conservative lower limits \citep{smith10gy}.

% Late time emission from SNe IIn (and what X-ray and radio mean)
At late times, the SN IIn forward and reverse shocks triggered by the ejecta-CSM interaction can be probed with X-ray and radio observations. Synchrotron emission arises in the forward shock, peaking at radio wavelengths, while thermal soft ($<$10 keV) and hard ($\ge$10 keV) X-ray emission arises in the heated plasma of the reverse and forward shocks, respectively, providing direct constraints on shock properties and other physical parameters not obtainable at optical wavelengths \citep{Chandra_2025}. Observations and modeling of the late-time multiwavelength emission from SNe IIn thus provide insight regarding the CSM structure and extent, and consequently, further measures of the progenitor mass-loss rate in the centuries leading up to core-collapse \citep{Chevalier2003, Chevalier2017}.

%Observations at these wavelengths are vital to constrain both the mass-loss rates as well as the dynamics of the shocks. 
In theory, every SN IIn should emit both radio and X-ray emission, given the high-density CSM universal to the subtype. However, not all objects show the expected levels of radio and X-ray emission due to a variety of potential factors, including Compton cooling/synchrotron cooling and/or optical depth effects at early times. It has also been suggested that there may be fundamental differences between those Type IIn that show multiwavelength emission and those that do not, perhaps relating to the structure of the CSM in addition to other unknown factors \citep{Chandra2018}. Up to this point, only around $\sim$20 SNe IIn have been detected at radio and X-ray wavelengths \citep{Chandra2018,Chandra_2025}. 

%Paper structure
We present X-ray and radio follow-up observations of four old (${\gtrsim}10$ years old) SNe IIn that exhibited prolonged and luminous optical and infrared (OIR) emission, indicating ongoing CSM interaction. The paper is laid out as follows: in Section \ref{subsec:targs} we introduce the sample and review previous studies of each system. In Section \ref{sec:Observations}, we describe the observations conducted for this study, and in Section \ref{sec:Analysis} we detail the analysis of the X-ray (\ref{subsec:X_an}) and radio (\ref{subsec:R_an}) observations. In Sections \ref{sec:Results} and \ref{sec:Discussion} we present our results and review our findings in the context of our current understanding of Type IIn supernovae. A summary of the work is presented in Section \ref{sec:Conclusion}.

\subsection{Targets} \label{subsec:targs}
In the following subsections, we present the four supernovae and review previous studies of each system. Source properties are summarized in Table \ref{table:properties}.

\begin{table*}[ht]
%\caption{Source Properties}
\centering
\begin{tabular}{l c c c l c c} 
\hline\hline %inserts double horizontal lines
Supernova ID & Host Galaxy & Redshift &Distance (Mpc) & Optical Detection Date & Peak \eq{ M_{g}} (mag) \\ [0.5ex] 
\hline 
SN 2013L$^{\alpha}$ & ESO 216-39 & 0.016992&72 & 2013 January 22 & $-19.15$ \\ 
SN 2014ab$^{\beta}$ & VV 306c & 0.02262 &106& 2014 January 12 & $-18.5$ \\ 
SN 2015da$^{\gamma}$ & NGC 5337 & 0.00667 & 53.2&2015 January 9 & $-20.18$ \\
KISS15s$^{\delta}$ & SDSS J030831.67-005008.6 & 0.03782 &156& 2015 July 31 & $-18.6$ \\ [1ex] 
\hline 
\end{tabular}
\centering
%\captionsetup{width=.95\textwidth}
\caption{Basic properties of the four supernovae observed in this study. Values for \eq{\alpha}, \eq{\beta}, \eq{\gamma}, \eq{\delta} adopted from \citet{Taddia2020}, \citet{Moriya2020}, \citet{Tartaglia2020}, and \citet{Kokubo2019}, respectively.}
%\caption*{Basic properties of the four supernovae observed in this study. Values for \eq{\alpha}, \eq{\beta}, \eq{\gamma}, \eq{\delta} adopted from \citet{Taddia2020}, \citet{Moriya2020}, \citet{Tartaglia2020}, and \citet{Kokubo2019}, respectively.}
% $^\alpha$\citet{Taddia2020} \\
% $^\beta$\citet{Moriya2020} \\
% $^\gamma$\citet{Tartaglia2020} \\
% $^\delta$\citet{Kokubo2019}
\label{table:properties}
\end{table*}

\subsubsection{SN 2013L} \label{2013L}
Supernova 2013L [RA(J2000) = $11^{\rm h}\,45^{\rm m}\,29.55^{\rm s}$, Dec(J2000) = $-50^\circ\,35'\,53.1''$] was discovered on 2013 Jan 22.025 UT in a southeastern arm of the galaxy ESO 216-39 with an unfiltered apparent magnitude of 15.6 \citep{Monard2013}. The supernova was discovered 19 days after the last non-detection \citep{Andrews2017}, so we assume the date of explosion $\approx$ the date of discovery given our observations are $>4000$ days later.  \citet{Andrews2017} conducted optical and infrared spectroscopic and photometric monitoring of SN 2013L for four years post-explosion. Given the slow evolution of asymmetric and multi-peaked \eq{H\alpha} and \eq{Pa\beta} emission lines, \citet{Andrews2017} suggest that SN 2013L is encased in a disk or torus of CSM, possibly arising from a massive interacting binary system. From the \eq{H\alpha} P-Cygni profiles, \citet{Andrews2017} measure a progenitor wind speed of \eq{v_{w}=80-130\,km\,s^{-1}} and a mass-loss rate of \eq{(0.3-8)\times10^{-3}\,M_{\sun}~yr^{-1}}, values consistent with a red supergiant or a yellow hypergiant progenitor star. 

\citet{Taddia2020} conducted ultra-violet (UV) to mid-infrared (MIR) photometry and spectroscopy of SN 2013L from +2 to +887 days post-discovery as part of the Carnegie Supernova Project II. They fit a two-component blackbody model to the spectral energy distributions ($>$132 days) to construct a bolometric light curve and infer a peak bolometric luminosity of \eq{\gtrsim3\times10^{43}\mathrm{erg~s}^{-1}} for SN 2013L. Similarly to \citet{Andrews2017}, \citet{Taddia2020} conclude that the CSM in SN 2013L is distributed anisotropically, but they calculate a higher wind velocity of \eq{v_{w}=120-240~km\,s^{-1}} and a progenitor mass-loss rate of \eq{(0.17-1.5)\times10^{-1}\,M_{\sun}~yr^{-1}}, both within the LBV regime. They also find an extreme infrared excess, suggesting potential dust formation.

\subsubsection{SN 2014ab} \label{2014ab}
Supernova 2014ab [RA(J2000) = $13^{\rm h}\,48^{\rm m}\,05.99^{\rm s}$, Dec(J2000) = $+07^\circ\,23'\,16.4''$] was discovered post-peak on 2014 Mar 9.43 UT with the Catalina Sky Survey (CSS) in the northern part of VV 306c, at an apparent V-band magnitude of 16.4 \citep{Howerton2014}. The supernova was retrospectively discovered to have appeared in the CSS 56 days earlier, on 2014 Jan 12, with an estimated explosion date of at least an additional month prior \citep{Bilinski2020}. The last non-detection was 212 days prior, but this only introduces an error $<10\%$ in any calculations involving the date of explosion, so we adopt the date of discovery as the date of explosion for simplicity.

\citet{Moriya2020} analyze early-time optical and NIR photometry and spectroscopy of SN 2014ab along with \eq{>}4 years of MIR WISE imaging. They measure a peak luminosity of \eq{>1\times10^{43}\mathrm{erg~s}^{-1}}. From the narrow H$\alpha$ P-cygni profile, they calculate a CSM speed of \eq{v_{w}=80\,km\,s^{-1}}. They estimate a mass-loss rate on the order of \eq{\dot M \sim 0.3\,M_{\sun}~yr^{-1}}, indicative of a high-mass progenitor such as an LBV star or possibly a binary system. \citet{Moriya2020} note excess infrared emission which they explain as radiative cooling of existing dust grains rather than formation of new dust, due to a lack of observed asymmetry in the blueshifted optical emission lines.

\citet{Bilinski2020} also conducted an early-time monitoring campaign of 2014ab, collecting optical spectroscopy, photometry, and spectropolarimetry. In contrast with \citet{Moriya2020}, \citet{Bilinski2020} identify asymmetrical blueshifted emission. The blueshift does not exhibit a strong wavelength dependency, so \citet{Bilinski2020} conclude that the asymmetric line profiles are likely the result of asymmetry in the CSM itself. The spectropolarimetric analysis suggests that SN 2014ab exhibits circular symmetry in the face-on plane. Thus, \citet{Bilinski2020} suggest that CSM geometry in SN 2014ab may be disk- or torus-like, with an inclination close to $0º$ (polar viewing angle). \citet{Bilinski2020} find a wind speed of \eq{v_{w}\sim80\,kms^{-1}} and mass-loss rate on the order of \eq{1M_{\sun}~yr^{-1}} for the SN 2014ab progenitor.

\subsubsection{SN 2015da} \label{2015da}
Supernova 2015da [RA(J2000) = $13^{\rm h}\,52^{\rm m}\,24.11^{\rm s}$, Dec(J2000) = $+39^\circ\,41'\,28.6''$] was discovered on 2015 Jan 9.90 UT in the northeastern part of the nearby spiral galaxy NGC 5337 \citep{Tartaglia2019}. The explosion date is the same as the discovery date as there was a deep upper limit at the SN location obtained 1.5 days prior to the discovery. \citet{Tartaglia2020} conducted more than four years of spectroscopic and photometric OIR monitoring of SN 2015da. SN 2015da shows similarities to SNe 2010jl and 2013L --- it is extremely luminous, with a peak bolometric luminosity of \eq{>3.0\times10^{43}\mathrm{erg~s}^{-1}} and a total radiated energy of \eq{E_{rad}>1.6\times10^{51}erg\,s^{-1}} over the first 2700 days \citep{Tartaglia2020, Smith2024}, exhibiting a slowly evolving optical continuum and excess infrared emission. From the evolution of their pseudo-bolometric lightcurves, \citet{Tartaglia2020} conclude that SN 2015da is encased in extended CSM produced by a progenitor undergoing eruptive mass-loss events with rates of \eq{0.6-0.7\,M_{\sun}~yr^{-1}}. 

\citet{Smith2024} carried out over 8 years of photometry and spectroscopy of SN 2015da and find similarly high mass-loss rates from the bolometric light curves (\eq{\dot M \sim0.04 - 0.1\,M_{\sun}yr^{-1}}, rising to \eq{\sim 0.4\,M_{\sun}yr^{-1}} in the decades prior to explosion). From their calculated steady wind speed of \eq{v_{w}=90\,km\,s^{-1}}, \citet{Smith2024} suggest a binary system as the SN 2015da progenitor. They invoke formation of new dust within the cold dense shell (CDS, in between the forward and reverse shocks) as the driving mechanism behind the infrared excess paired with wavelength-dependent blueshifted asymmetric line profiles. 

%SN 2015da was selected for late-time multiwavelength follow-up due to the slow evolution of its lightcurves, its bright peak luminosity, and the late-time IR excess.

\subsubsection{KISS15s} \label{KISS15s}
Supernova KISS15s [RA(J2000) = $03^{\rm h}\,08^{\rm m}\,31.64^{\rm s}$, Dec(J2000) = $-00^\circ\,50'\,05.55''$] was discovered in the low-mass, low-metallicity star-forming galaxy SDSS J030831.67-005008.6 with the Kiso Supernova Survey (KISS) on 2015 Sept 18.78 UT \citep{Kokubo2019}. \citet{Kokubo2019} retrospectively found the SN to have been picked up with multiple OIR surveys in the weeks prior, with an earliest discovery as PS15bva by the Pan-STARRS Survey for Transients on 2015 August 30 \citep{Kaiser2010} and an earliest detection date of 2015 July 31 by the SkyMapper automated telescope at Siding Spring Observatory in Australia \citep[SkyMapper Southern Survey;][]{Wolf2018}. The date of explosion was constrained by \citet{Kokubo2019} to be roughly 1 month earlier, but we again adopt the date of discovery as the explosion date for simplicity, with errors of $<1\%$.

\citet{Kokubo2019} conducted long-duration ($>800$ days) OIR broadband and spectroscopic monitoring of KISS15s. The slow evolution of the optical continuum out to \eq{\sim600} days indicates that the emission from KISS15s is powered by prolonged CSM interaction. Using an optical and IR synthesized pseudo-bolometric luminosity (\eq{L_{peak}\sim1.0\times10^{43}\mathrm{erg~s}^{-1}}) and an assumed CSM speed of \eq{v_{w}=40~km\,s^{-1}}, \citet{Kokubo2019} derive a progenitor mass-loss rate of \eq{\dot M \sim 0.4\,M_{\sun}~yr^{-1}}. While this mass-loss rate is indicative of episodic eruptions from an LBV star, it is highly dependent on observationally unconstrained factors. We note that LBV CSM speeds are expected to be higher than the assumed value of \eq{v_{w}=40~km\,s^{-1}}; they are typically \eq{\sim100-600~km\,s^{-1}} \citep{Smith2014}.

Due to the presence of blue-shifted components of the \eq{H\alpha} emission lines in the KISS15s spectra, \citet{Kokubo2019} propose that the region of ejecta-CSM interaction is asymmetric and inhomogeneous. The proposed system features a disk-like region of under-dense CSM (through which shocks would travel relatively quickly), encased in a more spherical dense CSM region. Additionally, while the optical emission from KISS15s decreased over the first \eq{\sim800} days post-explosion, the infrared continuum emission intensified, which \citet{Kokubo2019} interpret as the result of the formation of new dust within the CDS.

\section{Observations}\label{sec:Observations}
We obtained late-time X-ray (Section \ref{subsec:X_obs}) and radio (Section \ref{subsec:R_obs}) observations of our sample of four Type IIn supernovae. We describe these observations in the following subsections.

\subsection{X-ray}\label{subsec:X_obs}
\begin{figure}%
\centering
\begin{minipage}{.48\textwidth}
    \centering
    \includegraphics[width=.9\linewidth]{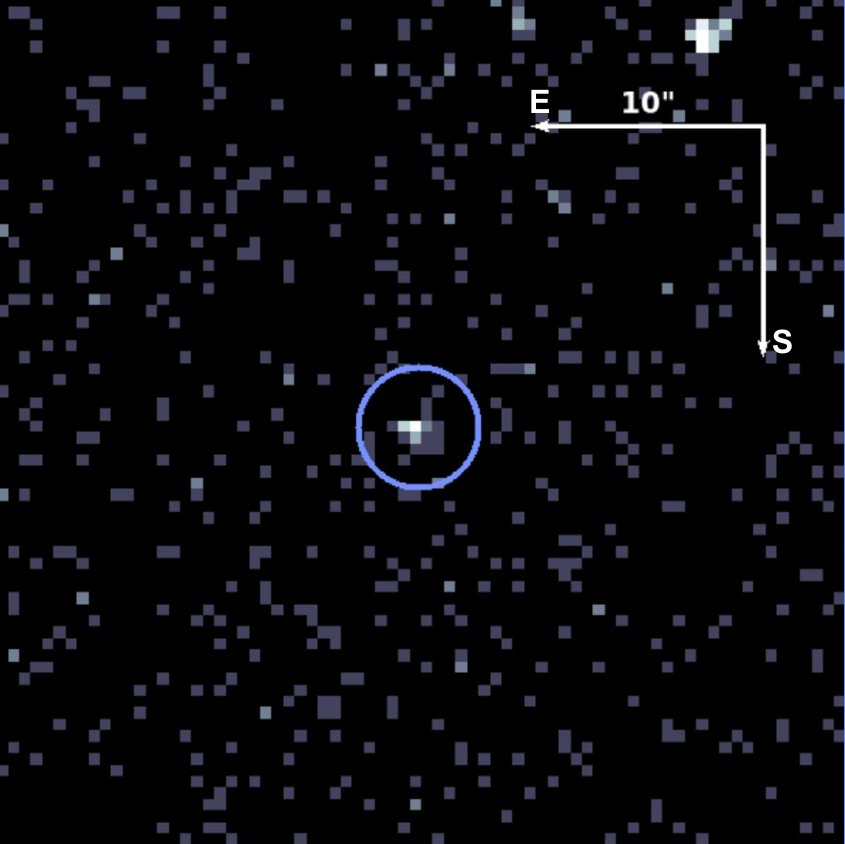}
    %\caption*{SN 2013L}
\end{minipage}%
\begin{minipage}{.48\textwidth}
    \centering
    \includegraphics[width=.9\linewidth]{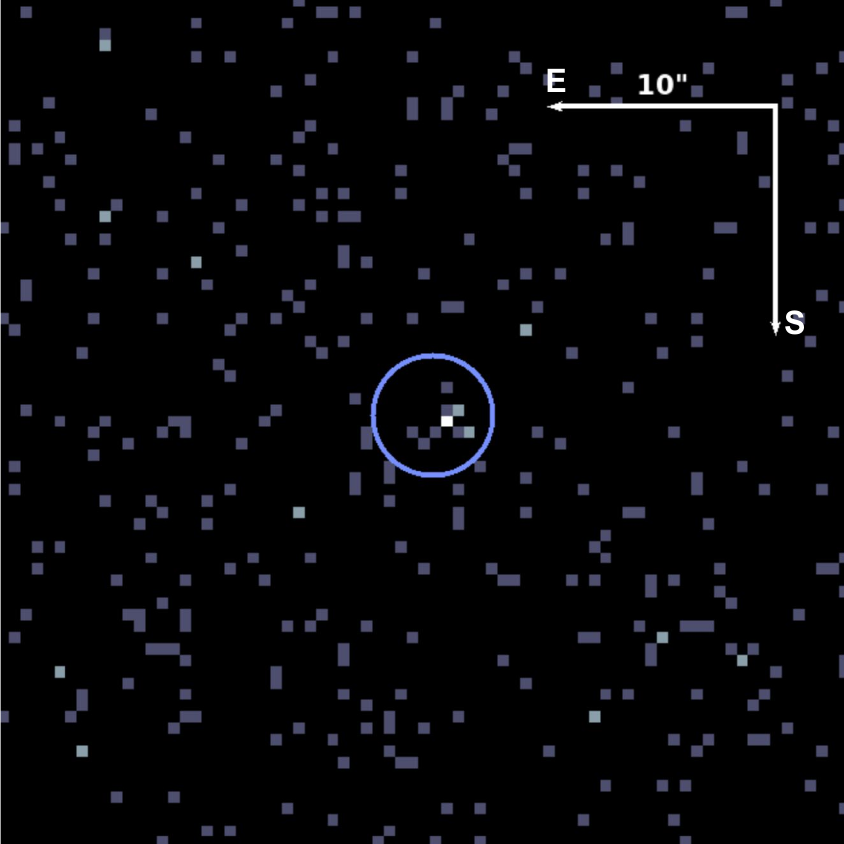}
    %\caption*{KISS15s}
\end{minipage}
\caption{Chandra ACIS-S (0.2-10\,keV) unsubtracted single epoch observations of SN 2013L (above) and KISS15s (below). SNe circled in blue.}
\label{fig:x_sources}
\end{figure}

All four supernovae were observed with the Chandra Advanced CCD Imaging Spectrometer (ACIS-S) instrument with no grating in VFAINT mode (proposal 25500415). As detailed in Table \ref{table:x_observations}, we obtained one exposure each of SNe 2014ab and 2015da, two exposures of SN 2013L, and four exposures of KISS15s. We detected X-ray emission from SN 2013L and KISS15s. Unsubtracted single exposure source images are shown in Figure \ref{fig:x_sources}. For the detected sources, we cross-referenced X-ray observations with the optical coordinates of the SN host galaxies and found that both host galaxies were X-ray faint. Other X-ray sources identified in our observations were found to be sufficiently far ($>$5") from the supernovae to avoid contaminating emission\footnote{\url{https://cxc.harvard.edu/proposer/POG/html/chap6.html\#tth_sEc6.6}}. Thus, the possibility of contamination from external sources was definitively ruled out. X-ray analysis is described in Section \ref{subsec:X_an}. 

\begin{table*}[ht]
%\caption{X-ray Observation Details}
\centering
\begin{tabular}{c c c c c} 
\hline\hline 
Instrument & Source & Observation Date & Exposure Time (ks) & Epoch (d)$^{\alpha}$ \\ 
[0.5ex] 
\hline 
Chandra ACIS-S & SN 2013L & 2024 Jul 2 & 10.08 & 4180 \\ 
Chandra ACIS-S & SN 2013L & 2024 Jul 3 & 42.13 & 4181 \\ 
Chandra ACIS-S & SN 2014ab & 2025 Jun 25 & 57.00 & 4183 \\
Chandra ACIS-S & SN 2015da & 2024 Jan 18 & 13.08 & 3297 \\
Chandra ACIS-S & KISS15s  & 2024 Nov 28 & 29.09 & 3409 \\
Chandra ACIS-S & KISS15s  & 2024 Dec 29 & 30.28 & 3440 \\
Chandra ACIS-S & KISS15s  & 2025 Jan 20 & 14.94 & 3462 \\
Chandra ACIS-S & KISS15s  & 2025 Jan 25 & 17.95 & 3467 \\
\hline
\end{tabular}
\centering
%\captionsetup{width=\textwidth}
\caption{X-ray observation details.  $^{\alpha}$Values represent lower limits (taken from earliest detection date).}
\label{table:x_observations}
\end{table*}

\subsection{Radio}\label{subsec:R_obs}

As cataloged in Table \ref{table:r_observations}, SNe 2014ab, 2015da, and KISS15s were observed with the Karl G. Jansky Very Large Array (VLA) (SN 2013L was too far south to be observed with the VLA) in K (18--26.5\,GHz), Ku (12--18\,GHz), X (8--12\,GHz), C (4--8\,GHz), and S (2--4\,GHz) bands, for a total observation time of 9 hours. Each observing block consisted of a flux calibrator, a phase calibrator, and science scans. 
%(calibration targets are listed in table \ref{table:calibrators}). 

All supernovae were observed with the Giant Metrewave Radio Telescope (GMRT). All four sources were observed in band 3 (250-500\,MHz), all but SN 2015da were observed in band 4 (550-850\,MHz), and SN 2013L was additionally observed in band 5 (1050-1460\,MHz). Each source was observed along with a flux calibrator and phase calibrator, accumulating to a total observation time of 22 hours. Imaging and analysis of VLA and GMRT observations are detailed in Section \ref{subsec:R_an}.  
%(see list of calibrators in table \ref{table:calibrators}). 

\begin{table*}[ht]
%\caption{Radio Observations Log}
\centering
\begin{tabular}{c c c c c c c} 
\hline\hline 
Telescope & SN & Obs Date & Epoch(d)$^{\alpha}$ & Central Frequency (GHz) & Flux Density (mJy) & Luminosity (\eq{erg\,s^{-1}\,Hz^{-1}})\\ 
[0.5ex] 
\hline
GMRT & SN 2013L  & 2024 Nov 11 & 4312 & 0.65  & $<$0.36      & $<2.23\times10^{27}$\\
GMRT & SN 2013L  & 2024 Nov 15 & 4315 & 0.4   & $<$1.00      & $<6.20\times10^{27}$\\
GMRT & SN 2013L  & 2024 Nov 19 & 4319 & 1.265 & $<$0.105     &  $<6.51\times10^{26}$\\
\hline
GMRT & SN 2014ab & 2025 Feb 7  & 4044 & 0.4   & $<$0.91      & $<1.22\times10^{28}$ \\
GMRT & SN 2014ab & 2025 Feb 5  & 4042 & 0.65  & $<$0.42      & $<5.65\times10^{27}$ \\
VLA  & SN 2014ab & 2025 Jun 24 & 4180 & 3     & $<$0.39      & $<5.24\times10^{27}$ \\
VLA  & SN 2014ab & 2025 Jun 24 & 4180 & 6     & $<$0.10      & $<1.34\times10^{27}$ \\
VLA  & SN 2014ab & 2025 Jun 24 & 4180 & 10    & $<$0.14      & $<1.88\times10^{27}$ \\
VLA  & SN 2014ab & 2025 Jun 26 & 4182 & 15    & $<$0.10      & $<1.34\times10^{27}$ \\
VLA  & SN 2014ab & 2025 Jun 26 & 4182 & 22    & $<$0.03      & $<4.03\times10^{26}$ \\
\hline

VLA  & SN 2015da & 2024 Feb 3  & 3313 & 3     & $<$0.37      & $<1.25\times10^{27}$ \\
VLA  & SN 2015da & 2024 Feb 3  & 3313 & 6     & $<$0.19      & $<6.43\times10^{26}$ \\
VLA  & SN 2015da & 2024 Feb 3  & 3313 & 10    & $<$0.17      & $<5.76\times10^{26}$ \\
VLA  & SN 2015da & 2024 Feb 14 & 3324 & 15    & $<$0.11      & $<3.73\times10^{26}$ \\
VLA  & SN 2015da & 2024 Feb 14 & 3324 & 22    & $<$0.17      & $<5.76\times10^{26}$ \\
GMRT & SN 2015da & 2024 Oct 28 & 3581 & 0.4   & $<$0.13      &  $<4.40\times10^{26}$\\
GMRT & SN 2015da & 2024 Oct 29 & 3581 & 0.65  & $<$0.11      &  $<3.73\times10^{26}$\\
\hline
VLA  & KISS15s  & 2024 Dec 3  & 3313 & 3 & 0.0627 ± 0.0159 & $(1.83\pm0.499)\times10^{27}$ \\
VLA  & KISS15s  & 2024 Dec 3  & 3313 & 6 & 0.0459 ± 0.00875  & $(1.34\pm0.275)\times10^{27}$ \\
VLA  & KISS15s  & 2024 Dec 3  & 3313 & 10 & 0.0367 ± 0.0129  & $(1.07\pm0.405)\times10^{27}$ \\
VLA  & KISS15s  & 2024 Dec 18 & 3324 & 15 & 0.0308 ± 0.0105  & $(8.97\pm3.30)\times10^{26}$ \\
VLA  & KISS15s  & 2024 Dec 18 & 3324 & 22 & 0.0463 ± 0.0138  & $(1.35\pm0.433)\times10^{27}$ \\
GMRT & KISS15s  & 2025 Feb 1  & 3490 & 0.4   & $<$0.23       & $<6.70\times10^{27}$\\
GMRT & KISS15s  & 2025 Jan 31 & 3489 & 0.65  & $<$0.11       & $<3.20\times10^{27}$\\
\hline

\end{tabular}
\centering
%\captionsetup{width=\textwidth}
\caption{Radio observation log. All upper limits are 3$\sigma$.  $^{\alpha}$Values represent lower limits (taken from optical discovery date).}
\label{table:r_observations}
\end{table*}

\needspace{2\baselineskip} % Fix standalone section title

\section{Data Analysis}\label{sec:Analysis}
We analyzed the radio and X-ray data in the standard manner for each telescope. To convert fluxes to luminosities, we used distances from Table \ref{table:properties} obtained from the optical papers on these objects. We describe our reductions in the following subsections. 

\subsection{X-Ray Analysis} \label{subsec:X_an}
We used the Chandra Interactive Analysis of Observations \citep[CIAO,][]{CIAO2006} software to reprocess all data, extract source and background spectra, and combine spectra for sources with multiple observations (see Table \ref{table:x_observations} for observation details). To maximize the encircled counts fraction based on the instrumental PSF\footnote{\url{https://cxc.cfa.harvard.edu/ciao/PSFs/psf_central.html}}, regions were standardized across all supernovae as a circle with radius of 2.5" for the source and a large annulus with an inner radius of 4" and an outer radius of 8-10" for the background (smaller for 2013L to avoid contamination), both centered on source coordinates.

We used the {\fontfamily{qcr}\selectfont CIAO} tool \texttt{aplimits} to calculate minimum count rates to constitute a source detection in each observation. For undetected sources, we found an upper limit on count rate from \texttt{aplimits} and input this into the \texttt{PIMMS}\footnote{\url{https://cxc.harvard.edu/toolkit/pimms.jsp}} observation planning tool (assuming a 1 keV reverse shock and line-of-sight galactic column density) to estimate the flux and luminosity upper limits of the supernovae as undetected Poisson sources, reported in Table \ref{table:x_parameters}. %and shown in Figure \ref{fig:luminosities}.
SN 2013L and KISS15s were both detected in all Chandra observations, while SN 2015da and SN 2014ab were not detected.

We conducted spectral analysis of the two detected supernovae using NASA's HEASARC software package {\fontfamily{qcr}\selectfont XANADU} \citep{HEASARC2014}. We binned each multi-observation source spectrum to a minimum of 5 counts/bin using \texttt{GRPPHA}. We then modeled the background-subtracted binned spectra using the fitting package \texttt{Xspec}. Given the low count rates for both detected sources, we used C-statistics (Poisson source with a Poisson background) as our fit statistic. We determined model goodness-of-fit with \eq{\chi^2}. 

Our models account for thermal plasma emission as well as line-of-sight absorption due to neutral hydrogen (hereafter referred to as column density, or \eq{N_{H}}). We tested two different plasma emission models: {\fontfamily{qcr}\selectfont bremss} and {\fontfamily{qcr}\selectfont apec}, selecting host galaxy metallicity measurements from previous literature. For SN 2013L, the {\fontfamily{qcr}\selectfont apec} model provided the optimal fit (as measured by \eq{\chi^2}), while the {\fontfamily{qcr}\selectfont bremss} model provided the more accurate fit for KISS15s.

\begin{figure}%
\centering
\begin{minipage}{.49\textwidth}
    \centering
    \includegraphics[width=\linewidth]{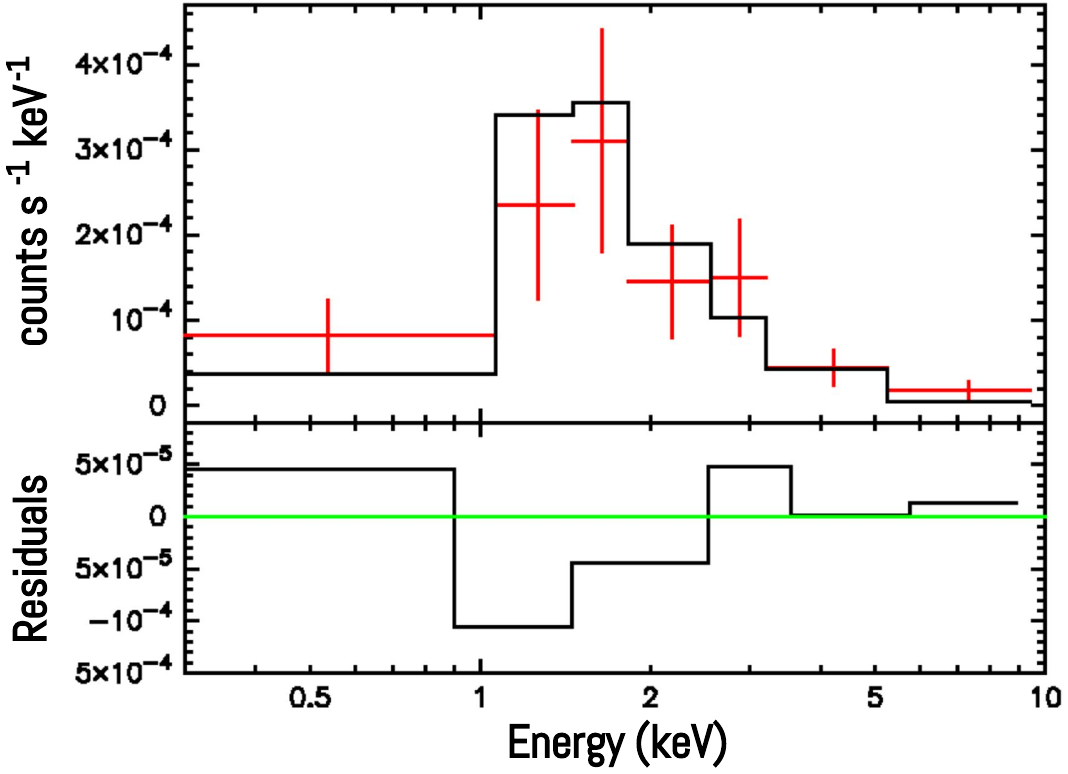}
    %\caption*{SN 2013L}
\end{minipage}%
\begin{minipage}{.5\textwidth}
    \centering
    \includegraphics[width=\linewidth]{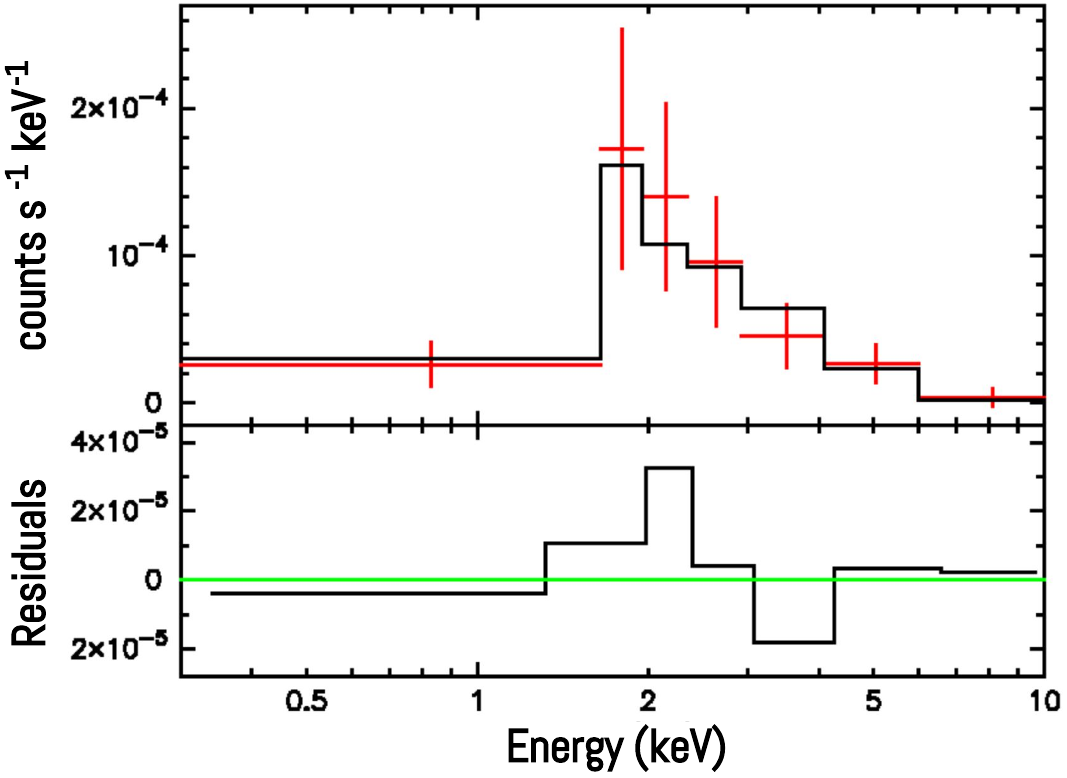}
    %\caption*{KISS15s}
\end{minipage}
\caption{Best-fit models for the two X-ray detected supernovae, SN 2013L (above) and KISS15s (below). We show the combined and binned source spectra (red--Chandra ACIS-S) with best-fit models and residuals in black. Parameters for the models are given in Table \ref{table:x_parameters}.}
\label{fig:fits}
\end{figure}

Fits were obtained by allowing all parameters to vary within physically motivated limits. The modeled plasma temperatures of both SN 2013L and KISS15s fell into the expected range for the reverse shock \citep[$\sim$1--5\,keV at late times;][]{Chevalier2017}, verifying expectations that the soft X-ray spectra from these old SNe would be dominated by reverse shock emission. In subsequent fits, we froze the temperature parameter at the initial best-fit value while allowing the normalization factor (representing flux) and column density to vary. This method increases degrees of freedom, stabilizing the fit and allowing for more reliable propagation of uncertainties within the low-count regime. Spectra with models and residuals are displayed in Figure \ref{fig:fits}. 

\begin{figure}%
\centering
\begin{minipage}{.49\textwidth}
    \centering
    \includegraphics[width=\linewidth]{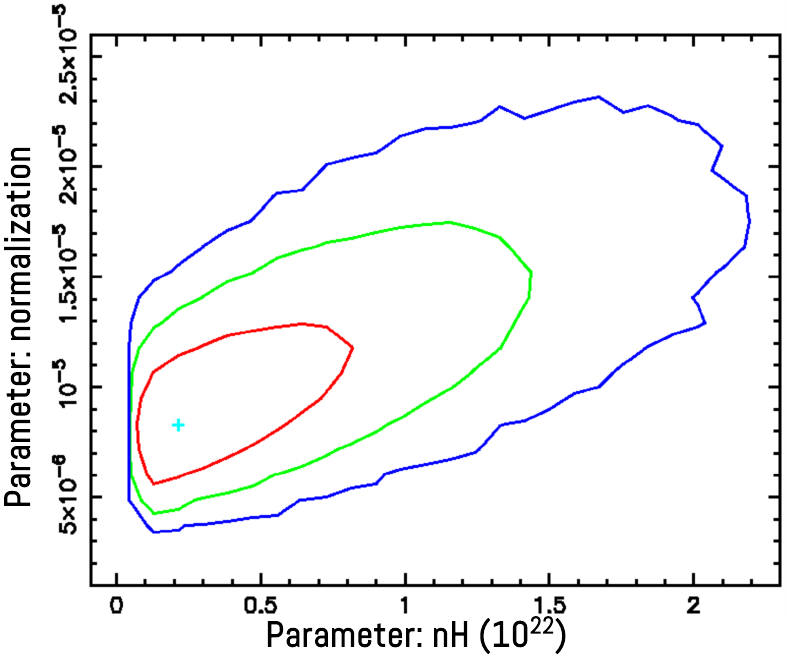}
    %\caption*{SN 2013L}
\end{minipage}%
\begin{minipage}{.49\textwidth}
    \centering
    \includegraphics[width=\linewidth]{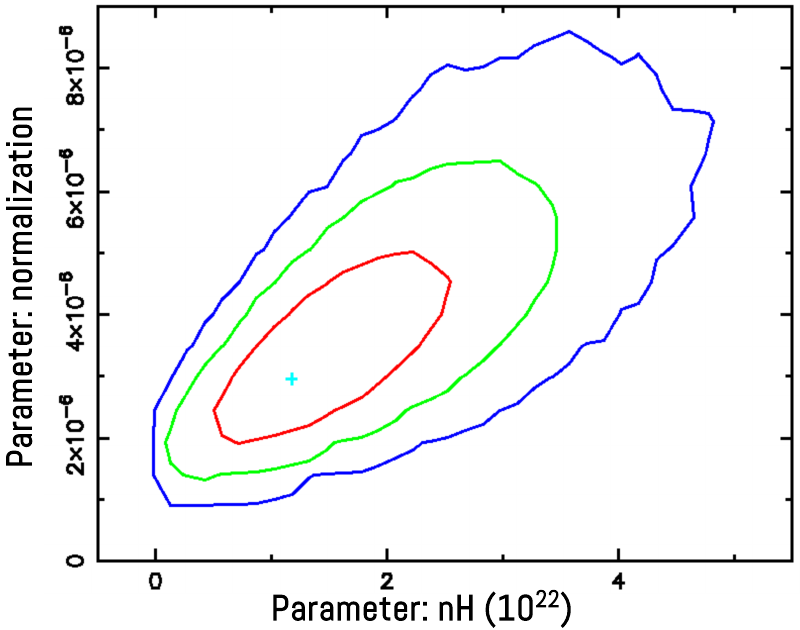}
    %\caption*{KISS15s}
\end{minipage}
\caption{Normalization and column density contour plots for SN 2013L (above) and KISS15s (below). \eq{1\sigma}, \eq{2\sigma}, and \eq{3\sigma} confidence regions are shown in red, green, and blue, respectively. Best-fit model values are denoted by the cyan cross. Flattening of curves at low \eq{\mathrm{N_{H}}} values signifies lower limit due to line-of-sight galactic hydrogen.}
\label{fig:contour_plots}
\end{figure}

\begin{figure}%
    \centering
    \includegraphics[width=.95\linewidth]{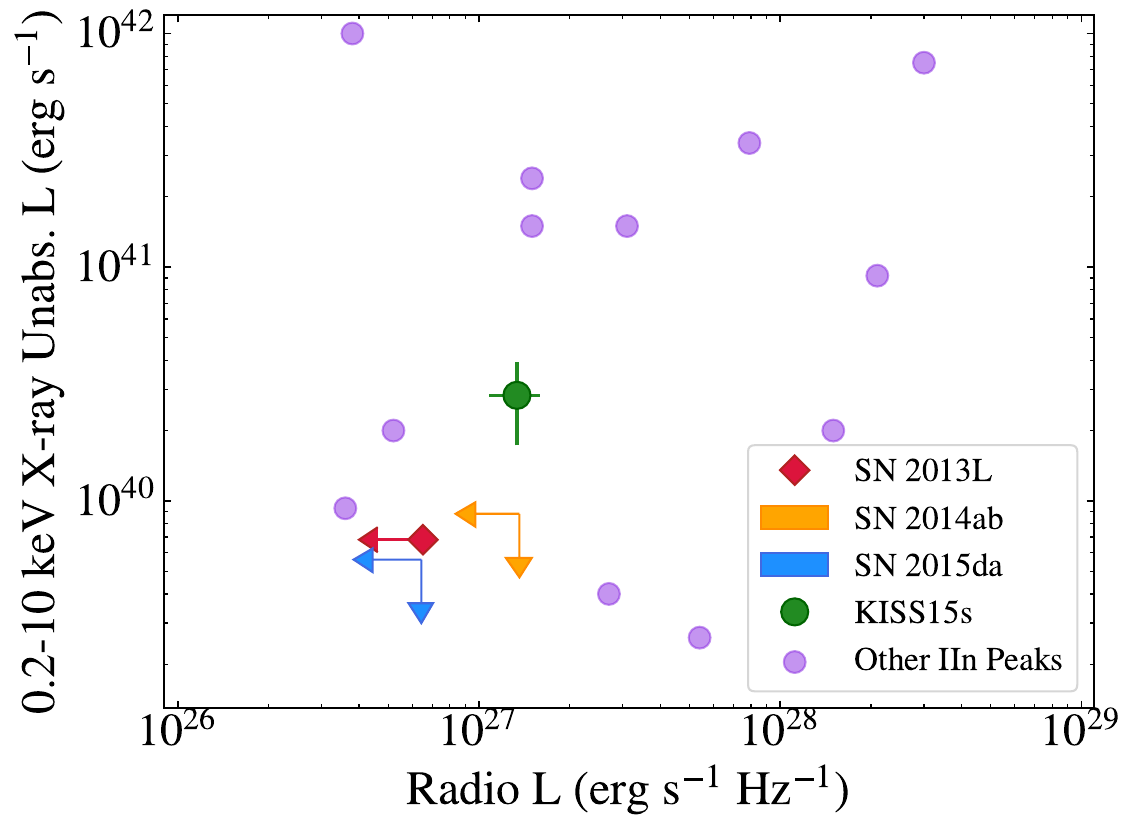}
    \caption{X-ray (0.2-10 keV) and radio luminosities of the four supernovae in the sample. For SN 2013L, we use the 1.25 GHz luminosity as it was not observed with the VLA. For all other SNe, we use the 6 GHz luminosity. We plot peak luminosity values of other SNe IIn from \cite{Chandra_2025}.}
    \label{fig:luminosities}
\end{figure}

Initial errors were estimated using \texttt{Xspec}'s {\fontfamily{qcr}\selectfont steppar}. We employed Goodman-Weare Markov Chain Monte Carlo (MCMC) sampling to map the parameter space and further constrain uncertainties, using a 1,000,000 step chain with 5,000 step burn-in and 20 walkers. We measured the column density, normalization, observed (absorbed) and intrinsic (unabsorbed) flux of the best fits, with errors derived from the chains. Fit and parameter details for each supernova are listed in Table \ref{table:x_parameters}. \eq{1\sigma}, \eq{2\sigma}, and \eq{3\sigma} confidence contour plots are displayed in Figure \ref{fig:contour_plots}. We calculated luminosities using the intrinsic flux and the distances reported in Table \ref{table:properties}. Unabsorbed luminosities are shown in Figure \ref{fig:luminosities}.

%X_ray parameters table 
\begin{table*}[ht]
%\caption{X-ray Modeling Parameters}
\centering
\begin{tabular}{ C c c c c c c c} 
\hline\hline 
$\rm{SN}$ & Ct. Rate (cts\,s$^{-1}$) & Model & $\chi^2_{\nu}$ & \eq{N_{H}} (\eq{10^{22}cm^{-2}}) & T (keV) & Unabs. $\mathrm{Flux (erg\, cm^{-2}s^{-1})}$ & Unabs. $\mathrm{L \,(erg \,s^{-1})}$\\  
\hline 
$\rm{2013L}$ &$6.901 \times 10^{-4}$ & apec * phabs & 0.86 & $0.24_{-0.14}^{+1.04}$ & 2.85 & $1.11^{+0.77}_{-0.45}\times 10^{-14}$& $6.89^{+1.31}_{-1.21} \times 10^{39}$ \\ 
$\rm{KISS15S}$ &$2.881 \times 10^{-4}$& bremss * phabs & 0.90 & $1.32_{-0.34}^{+1.05}$ & 3.20 & $0.97^{+0.38}_{-0.43}\times 10^{-14}$ & $2.83^{+1.60}_{-1.65} \times 10^{40}$\\
$\rm{2014ab}$ & $< 2.51 \times 10^{-4}$& N/A &... &... & ... & $< 6.55\times 10^{-15}$&$<8.81 \times 10^{39}$\\
$\rm{2015da}$& $< 6.97 \times 10^{-4}$ & N/A & ...&...&... & $< 1.66\times 10^{-14}$&$<5.61 \times 10^{39}$\\
\hline 
\end{tabular}
\centering
%\captionsetup{width=\textwidth}
\caption{Parameters from best-fit X-ray models, with flux and luminosity measured from 0.2-10 keV. The column densities we report are affected by absorption from the host galaxy itself. The reported temperatures are frozen to allow for more robust fits given the low number of datapoints. Note: for the SN 2013L column density, the posterior was truncated due to proximity of the best-fit value to the lower limit set by galactic \eq{N_H}, so we were unable to estimate a lower uncertainty. We thus report the value pertaining to the galactic column density towards SN 2013L (i.e. a CSM column density of 0) for the lower uncertainty of this parameter.}
\label{table:x_parameters}
\end{table*}

\needspace{2\baselineskip} % Fix standalone section title

\subsection{Radio Analysis} 
\label{subsec:R_an}

VLA observations were processed via a combination of manual and automated flagging and calibration \citep[VLA data calibration pipeline, Common Astronomy Software Applications (CASA) v6.6.1-17, ][]{McMullin2007}. In cases with both manual and pipeline calibrations, we cross-checked results (i.e. measured root mean square (RMS) noise and source flux) and found them to be consistent. GMRT observations of all four supernovae were reduced using the automated GMRT data reduction pipeline, CASA Pipeline-cumulative-Toolkit for Upgraded Giant Metrewave Radio Telescope data REduction (CAPTURE) with CASA v6.6.1-17 \citep{Kale2021}. Following initial data reduction, we performed self-calibration when needed to minimize sidelobe deconvolution errors and improve the final image quality. Images from the VLA and GMRT were processed using the CASA task {\fontfamily{qcr}\selectfont tclean} in interactive mode, with the clean threshold determined by \eq{2\times \mathrm{RMS}} of a large region near the source coordinates in the dirty image.

We conducted image analysis and fitting with the Cube Analysis and Rendering Tool for Astronomy \citep[CARTA,][]{CARTA2021}. 

SN 2014ab was a complex case as it is only offset from the host galaxy center by 1.5", and the galaxy is radio-bright. However, careful analysis of prior C band data of the SN (VLA Proposal ID 23A-328) revealed that the emission detected in the region is likely due to the host galaxy itself, given that the flux stays within 5$\%$ of its 2023 value across this two-year window. We therefore conclude that SN 2014ab is not detected at the current epoch. We additionally re-analyzed 2015 data of the SN at higher resolution (ID 15A-129), and found a weak detection (along with host galaxy flux density in line with the 2023/2025 values).  

\begin{figure}
    \centering
    \includegraphics[width=8 cm, height= 6 cm]{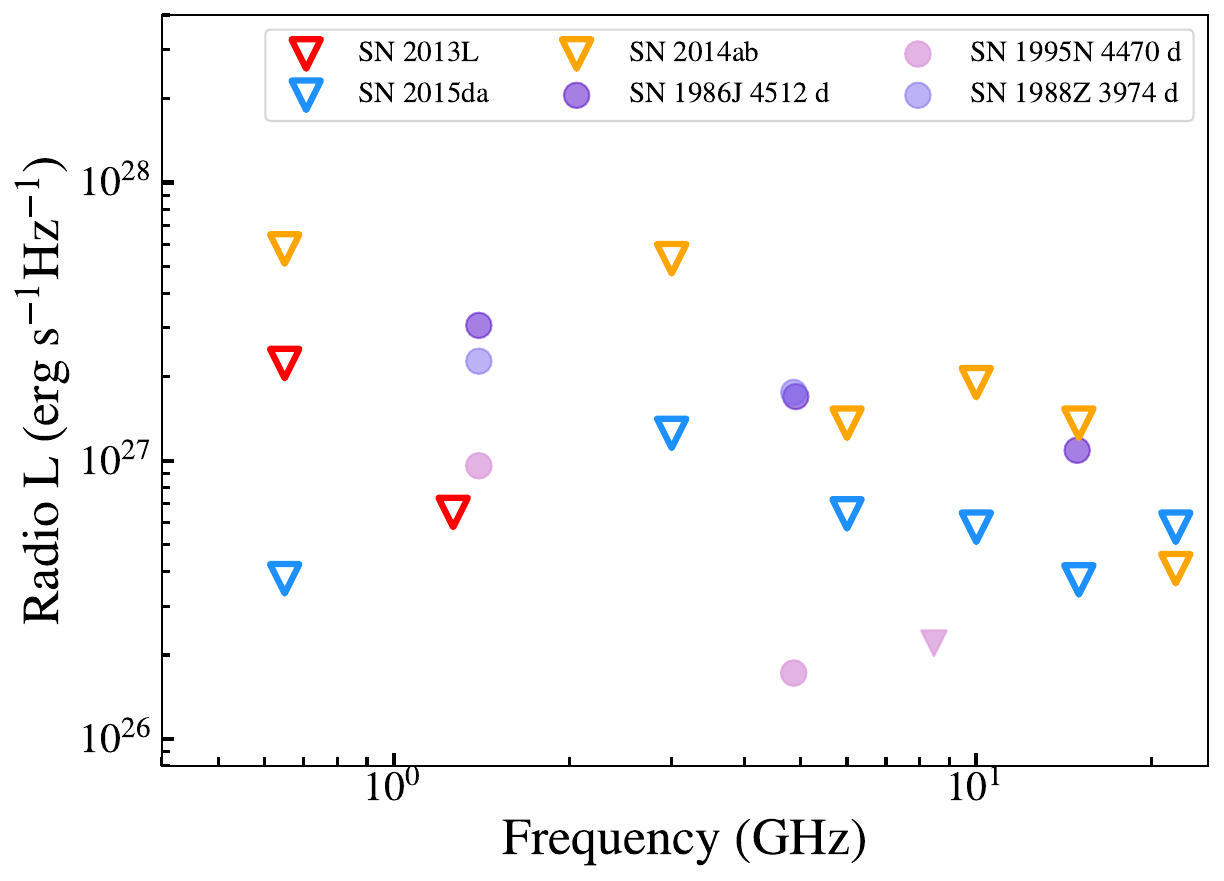}
    \caption{A view of some of the radio upper limits on the three non-detected supernovae, with detections at similarly very late times for other SNe IIn 1986J, 1988Z, and 1995N shown in shades of purple, from \cite{Bietenholz2010,  Williams_2002, Chandra_2009}. Details for the sources presented in this work are shown in Table \ref{table:r_observations}.}
    \label{fig:GMRT}
\end{figure}

SNe 2013L and 2015da were also not detected in any radio observations. We calculated upper limits for these sources using the RMS in a region several times the beam size centered on the source location. A statistically significant measurement is represented by a \eq{3\sigma} detection; thus, flux and luminosity upper limits were determined using \eq{3\times\mathrm{RMS}} (reported in Figures \ref{fig:luminosities} and \ref{fig:GMRT}, and Table \ref{table:r_observations}).

\begin{figure}%
\centering
\begin{minipage}{.156\textwidth}
    \centering
    \includegraphics[width=\linewidth]{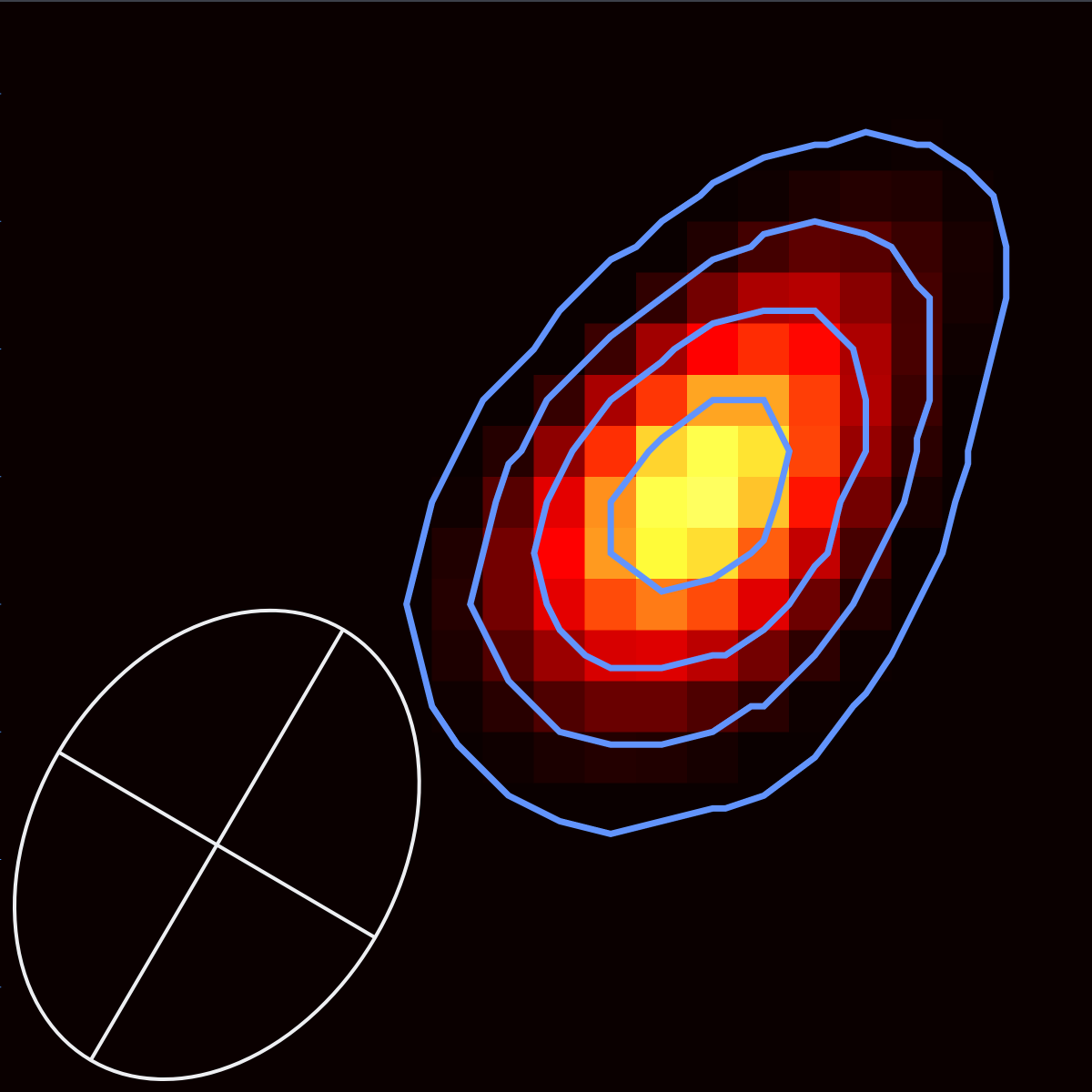}
    %\caption*{C Band}
\end{minipage}%
\hspace{-2 pt}\begin{minipage}{.156\textwidth}
    \centering
    \includegraphics[width=\linewidth]{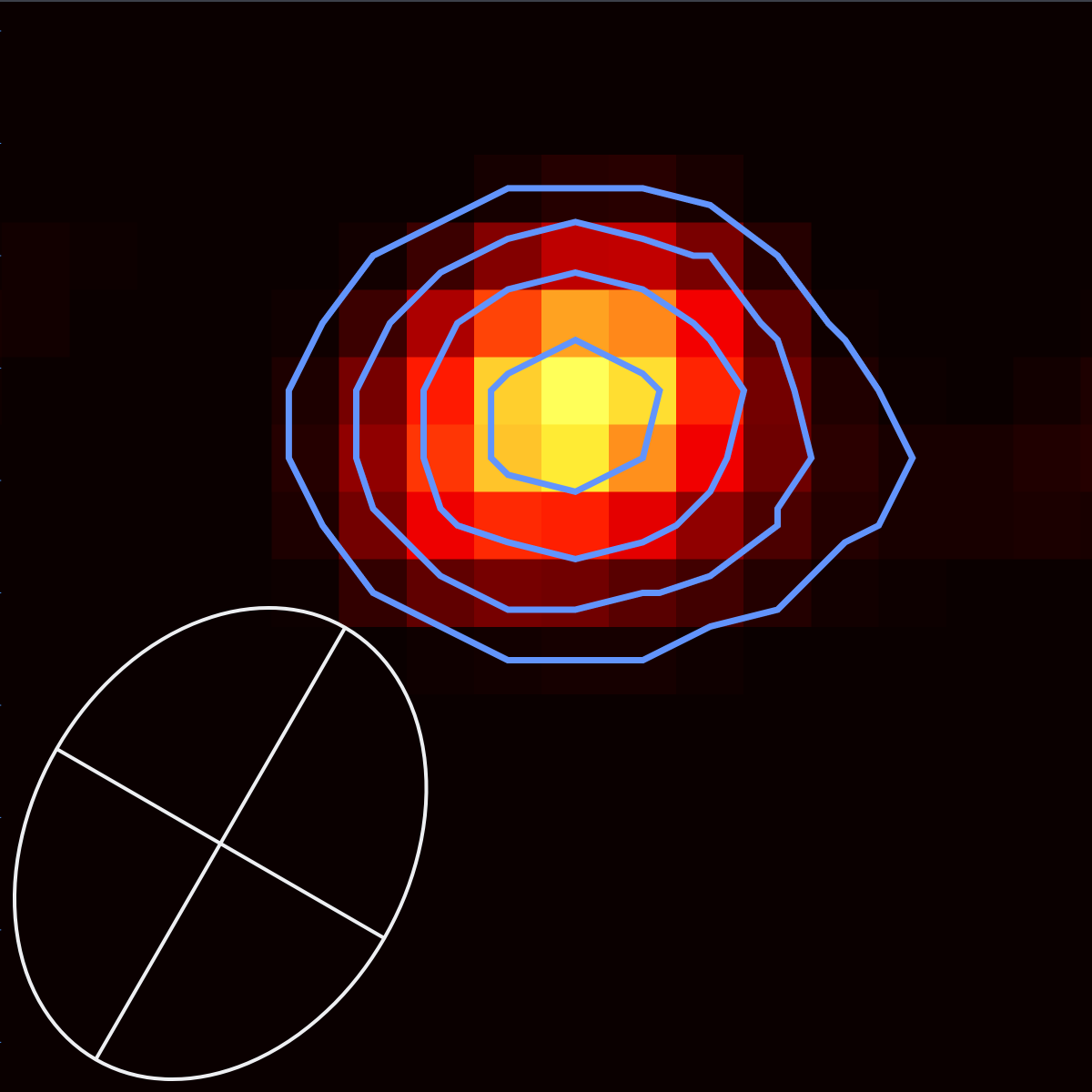}
    %\caption*{X Band}
\end{minipage}%
\hspace{-2 pt}\begin{minipage}{.156\textwidth}
    \centering
    \includegraphics[width=\linewidth]{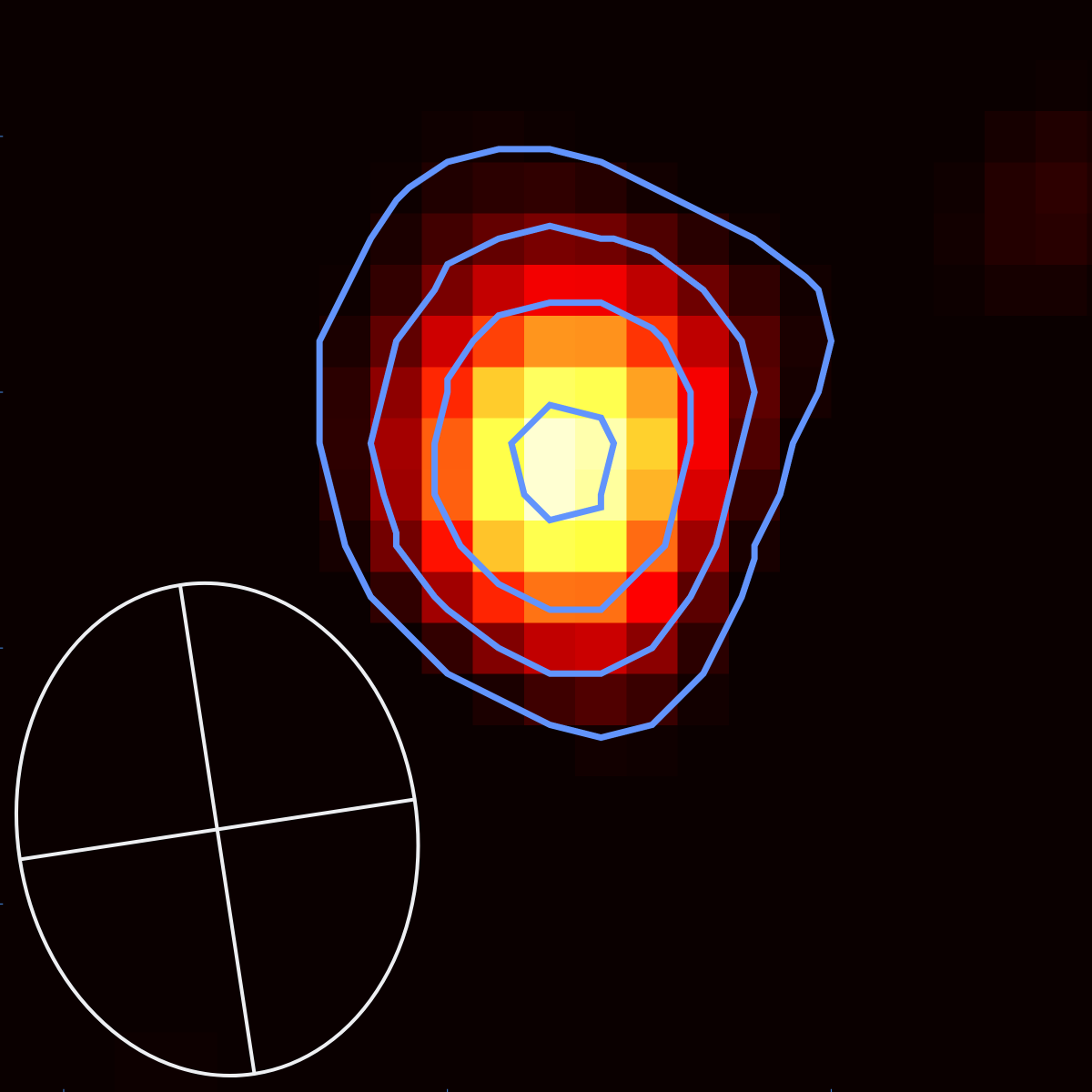}
    %\caption*{K Band}
\end{minipage}
\caption{KISS15s in VLA C band (4-8GHz, left), X band (8-12GHz, center), and K band (18-26.5GHz, right). The ellipse in the bottom left of each image signifies the beamsize (resolution element size). Contours are scaled individually for each band --- for C band, contours appear at 10, 21, 32, and 43 $\mu$Jy/beam; for X band, contours appear at 13, 18.5, 24, and 29.5 $\mu$Jy/beam; and for K band, contours appear at 18, 24, 30, and 36 $\mu$Jy/beam.}
\label{fig:r_im_K15s}
\end{figure}

We detected radio emission from KISS15s in all VLA bands (see Figure \ref{fig:r_im_K15s} for selected observations). We used the CARTA Image Fitting tool to fit 2D Gaussian models to the source emission in each band, from which we measured the flux density (from the integrated flux) and calculated corresponding luminosities. The radio spectrum of KISS15s is shown in Figure \ref{fig:k15s_r_fits}, and the results from fitting are detailed in Table \ref{table:r_observations}.

%Radio
% We produced spectral energy distributions of the radio-bright sources from the bandpass flux densities, measured as described in section \ref{subsec:R_an}. The spectral rise due to synchrotron emission is expected to follow a powerlaw, 
% \[S(\nu)\propto\nu^{-\alpha}\]
% Where S is flux density, $\nu$ is frequency, and $\alpha$ is the spectral index. The spectral index for synchrotron emission from SNe IIn in the optically thin regime is expected to be $\gtrsim0.5$ \citep[e.g.][]{Chevalier2017, Ellison1991}. 

\section{Results}\label{sec:Results}

The 0.2-10 keV X-ray and C band radio (except for 2013L, for which we report the L band upper limit) luminosities/upper limits of the sample are shown in Figure \ref{fig:luminosities}. We use the observed luminosities and non-detections to place constraints on the progenitor systems of each supernova.

\subsection{SN 2013L Results}
SN 2013L shows clear evidence for soft X-ray emission from the best-fit temperature of T=2.85 keV. The best-fit column density for SN 2013L is consistent with the value expected from galactic absorption \citep{Guver_2009} of $10^{21}\,\rm cm^{-2}$, indicating that the system may be coming to the end of its ejecta-CSM interaction phase. However, if we are viewing CSM that is non-spherical, it is possible that the absorption is missed due to viewing angle effects, such as if the CSM is denser at the equator but we are viewing closer to the poles.

To better understand the progenitor of SN 2013L, we first estimate the phase of the progenitor star's evolution that our X-ray observations probe at time $\rm{t_{obs}}$ using

\begin{equation}
\mathrm{t_{pre~SN} = t_{obs}\left(\frac{v_{s}}{v_w}\right)}
\label{eq:tpreSN}
\end{equation}

where $\mathrm{t_{pre~SN}}$ is the pre-explosion phase, $v_{s}$ is the forward shcok speed, and $v_w$ is the wind velocity. Using the observed wind velocity of \eq{v_w\sim 120~km\,s^{-1}} at early times (the blue velocity at zero intensity of the \eq{H\alpha} P cygni profile), a rough shock speed of $\sim$\eq{4000~km\,s^{-1}} estimated by \citet{Taddia2020}, and a weighted average age at observation of \eq{t_{obs}=4180.8d}, we estimate \eq{t_{pre~SN} \approx 382 yr}.
 
To estimate a mass-loss rate, we use the measured X-ray luminosity. We assume that the X-ray emission is coming from an adiabatic reverse shock at this late epoch. At thousands of days post-explosion, this should be the case for any Type IIn as they all have $10^{-8} \lesssim \rm{\dot M} \lesssim 10^{0}~\rm{M_{\odot}~yr^{-1}}$ (using equation 15 from \cite{Chevalier2017}). To verify this assumption, we use a mass-loss rate of $10^{-2}~\rm{M_{\odot}~yr^{-1}}$ and an average shock speed (over the course of the SN evolution) \eq{v_{\rm sh}=4500\,km~s} to calculate that both shocks are adiabatic at $>$2500 days post-explosion \citep{Chevalier2017}. With both shocks adiabatic, the forward shock will contribute $<5\%$ of the 1 keV luminosity (in the region where the majority of the emission is seen) no matter what CSM/ejecta density gradients $s/n$ are assumed \citep{Fransson_1996}. It is thus reasonable to assume that the forward shock does not contribute significantly to the luminosity and we are seeing emission predominatly from the reverse shock, especially considering that the \eq{\dot{M}} for 3 of our four objects is closer to $10^{-3}~\rm{M_{\odot}~yr^{-1}}$, which would only decrease this fraction (see later sections).

We then use formula 3.10 from \citet{Fransson_1996}, which gives the reverse shock adiabatic luminosity from steady mass-loss, setting the respective CSM and ejecta density profile exponents to $s=2$ and $n=7$ (with no means to constrain $n$ or $s$ for any of our objects with lacking data but using a shallower $n$ as has been seen in SNe IIn \citep{BaerWay2025}):
\begin{equation}
\begin{split}
L_{\rm rev}(1\,\mathrm{keV}) &=
1.39 \times 10^{39}\,
T_{8}^{-0.24}
\exp\!\left(-\frac{0.0116}{T_{8}}\right)
\,\xi\, C_{*}^{2} V_{4}^{-1} \\
&\quad \times
\left( \frac{t_{d}}{11.57} \right)^{-1}
\end{split}
\label{eq:Mdot}
\end{equation}
where $T_{8}$ is the temperature of the reverse shock normalized to \eq{10^8\,K}, $\xi$ represents the hydrogen abundance fraction (set to 0.86 for solar abundance), $C_{*}=\frac{\dot M_{-5}}{v_{w,10}}$ is the mass-loss rate normalized to $\mathrm{10^{-5}\,M_{\odot}~yr^{-1}}$ over the wind velocity normalized to 10\eq{~km\,s^{-1}}, \eq{V_{4}} is the reverse shock speed (in the frame of the observer) in units of \eq{10^4~km\,s^{-1}}, and $t_{d}$ is the time since explosion.

We use our luminosity, the rough ejecta speed, the 1 keV spectral luminosity from PIMMS (taken from the total 0.2 - 10 keV luminosity), and the fitted temperature of 2.85 keV to find a mass-loss rate (with a 120 km s$^{-1}$ wind speed from \citet{Taddia2013}) of $(2.0 \pm 1.3) \times 10^{-3}~\rm{M_{\odot}~yr^{-1}}$ (incorporating errors from the discrepant wind speeds found by \cite{Taddia2020} and \cite{Andrews2017}). This result is interpreted in Section \ref{sec:Discussion}.

SN 2013L was not detected in GMRT radio observations. To obtain a limit on the mass-loss rate, we model the 1.25 GHz (as this is the most constraining point) non-detection in a synchrotron scenario detailed by \citet{Chevalier_98} with the emission attenuated by both synchrotron self-absorption (SSA) and free-free absorption (FFA), with the free-free optical depth detailed by i.e. \citet{Weiler2002}. We assume for a Type IIn that the optically thin spectral index $\alpha=0.75$ \citep[][see Equation \ref{eq:powerlaw}]{Stroh_2021} and assume the CSM density profile exponent $s=2$ without other constraints. We take a shock speed \eq{\sim4000~km\,s^{-1}} (assuming the shock radius=\eq{v_{s}t}) and find that \eq{\dot{M}<4 \times 10^{-3}\,M_{\odot}~yr^{-1}} (considering the limit from a scenario where pure FFA is causing absorption at 1.25 GHz of $> 10\,\rm{M_{\odot}~yr^{-1}}$ as unphysically high based on the X-ray result). We thus find that the radio non-detection is consistent with the mass-loss rate inferred from the X-ray detection. A mass loss rate on the order of \eq{10^{-3}\, M_{\odot}~yr^{-1}} suggests a decreasing or possibly relatively constant mass-loss rate from early-time measurements on the order of \eq{10^{-3} - 10^{-1}\,M_\sun~yr^{-1}} measured by \citet{Andrews2017} and \citet{Taddia2020}, indicating that any enhanced mass-loss occurred mainly in the 400 years prior to explosion.

\subsection{SN 2014ab Results}
SN 2014ab was not detected at radio or X-ray wavelengths. We obtain limits on the radio and X-ray luminosity as shown in Figure \ref{fig:luminosities} based on the distance reported by \citet{Bilinski2020}. We use Equation \ref{eq:Mdot}, again taking the 1 keV spectral luminosity from the overall unabsorbed luminosity using PIMMS, with a 2000 km\,s$^{-1}$ shock speed based on the measured FWHM in optical profiles (the derived mass-loss is not changed by more than a factor 1.5 even when changing this value by a factor of 2) and the 80 \eq{km\,s^{-1}} wind speed from \citet{Bilinski2020} to derive \eq{\dot{M}<2 \times 10^{-3}\,\rm{M_{\odot}~yr^{-1}}}. We also find a radio limit on the mass-loss rate using the 3 GHz non-detection in the same manner as for SN 2013L of $<2.0 \times 10^{-2}\,\rm{M_{\odot}~yr^{-1}}$, a conservative estimate due to the relatively shallow upper limit. These limits suggest intensifying mass-loss in the centuries pre-explosion, given that \citet{Bilinski2020} measured a mass-loss rate of $\sim 1\, \rm{M_{\odot}\,yr^{-1}}$ (with \citet{Moriya2020} estimating a slightly lower $\sim 0.3\, \rm{M_{\odot}\,yr^{-1}}$) in the final years pre-explosion. If the shock has moved at around \eq{2000~km\,s^{-1}} on average throughout the SN evolution, our limit would correspond to mass-loss $\sim$300 years pre-explosion, indicating again only a few centuries of enhanced mass loss. 

\subsection{SN 2015da Results}
SN 2015da was not visible in soft X-ray or radio bands. We estimate flux and luminosity upper limits for both the 0.2 -- 10 keV X-ray and individual radio bands, shown in Figure \ref{fig:luminosities}. As with SN 2014ab, the X-ray and radio non-detections signify that the ejecta-CSM interaction in SN 2015da is likely coming to an end, or possibly finished altogether. \citet{Smith2024} find the intermediate width component of \eq{H\alpha} to have \eq{v_{FWHM}\sim2,400~km\,s^{-1}} on day 3,030. Using this value as an estimate of the forward shock velocity and the P cygni--constrained wind speed of \eq{v_{w}\sim90~km\,s^{-1}} from \citet{Smith2024}, we solve Equation \ref{eq:tpreSN} to find that the progenitor mass-loss was likely constrained to \eq{\lesssim 250} years pre-supernova, given the X-ray non-detection at 3297 days post-discovery. 

To obtain a limit on the mass-loss rate, we assume the X-ray emission at this last epoch originates from a 1 keV reverse shock. Using the same approach as for SN 2014ab with the same assumptions given the similar age, we find \eq{\dot{M}<1.7\times 10^{-3} \,M_{\odot}~yr^{-1}}. The radio limit on the mass-loss rate from the GMRT observations using the same assumptions as for the other SNe is  $<6 \times 10^{-3}\,\rm{M_{\odot}\,yr^{-1}}$. The X-ray limit is inconsistent with the extrapolated rate from \citet{Smith2024}, which predicts a rate \eq{\sim 10^{-2}\,M_{\odot}~yr^{-1}} at late phases $\sim$ 3000 days post-explosion. This suggests either an abrupt increase in mass-loss leading up to the supernova as the X-ray observations are at a slightly later phase ($\sim$ 300 days later than the last optical observations in \citet{Smith2024}), or some other effect such as asymmetry or clumping in the CSM. We elaborate in Section \ref{sec:Discussion}.

\subsection{KISS15s Results}
KISS15s is detected in both X-ray and radio bands, as described in Section \ref{sec:Analysis}, indicating ongoing ejecta-CSM interaction. First, we use the X-ray luminosity to calculate the mass-loss rate of the KISS15s progenitor in a similar manner as for SN 2013L.

With the inferred wind velocity of \eq{v_{w}\sim40~km\,s^{-1}} from \citet{Kokubo2019} (from \eq{v_{FWHM}} of the narrow emission lines), the estimated forward shock velocity \eq{\sim2000~km\,s^{-1}}, and the weighted average age of KISS15s during Chandra observations (3439d), we again use Equation \ref{eq:tpreSN} to estimate that the reverse shock is probing at least \eq{\sim450} years pre-supernova. 

Employing Equation \ref{eq:Mdot} with the ejecta speed and epoch of our observations, we estimate a mass-loss rate of \eq{\dot M \approx (4.0 \pm 1.6) \times 10^{-3} \,M_{\sun}~yr^{-1}} at \eq{\sim 450} years pre-explosion, a rate two orders of magnitude below the estimate of \citet{Kokubo2019}.

%Radio
The radio flux densities of KISS15s can also help constrain the mass-loss rate of the progenitor. Following \citet{Weiler2002}, the mass-loss rate at the peak of the radio synchrotron emission can be found by: 

\begin{align}
\rm{\dot{M}} \approx\; & (1\times10^{-6}) \left(\frac{\rm{{v_{w}}}}{10~\text{km}~\text{s}^{-1}}\right) 
\left( \frac{\rm{L_{6\text{cm peak}}}}{10^{26}~\text{erg}~\text{s}^{-1}~\text{Hz}^{-1}}\right)^{0.54} \notag \\
& \times \left( \frac{\rm{t_{6\text{cm peak}}}}{\text{d}}\right)^{0.38}
\left(\frac{\rm{M_{\odot}}}{\text{yr}} \right)~.
\label{eq:R_Mdot}
\end{align}

As these are the only radio observations of this supernova, and peak radio luminosities in SNe IIn are expected to occur at earlier times \citep{Chevalier2003}, we do not have measurements of the peak radio luminosity. We treat the measured C-band luminosity as a lower limit on the peak 6cm luminosity, and calculate a lower limit on the mass-loss rate of \eq{\dot M \gtrsim 3.85 \times 10^{-4}\,M_{\sun}~yr^{-1}}. The radio and X-ray detections indicate that, despite the decrease in mass-loss rate, the ejecta-CSM interaction is ongoing.

\begin{figure}
\centering
%\begin{minipage}{.48\textwidth}
    %\centering
\includegraphics[width=\linewidth]{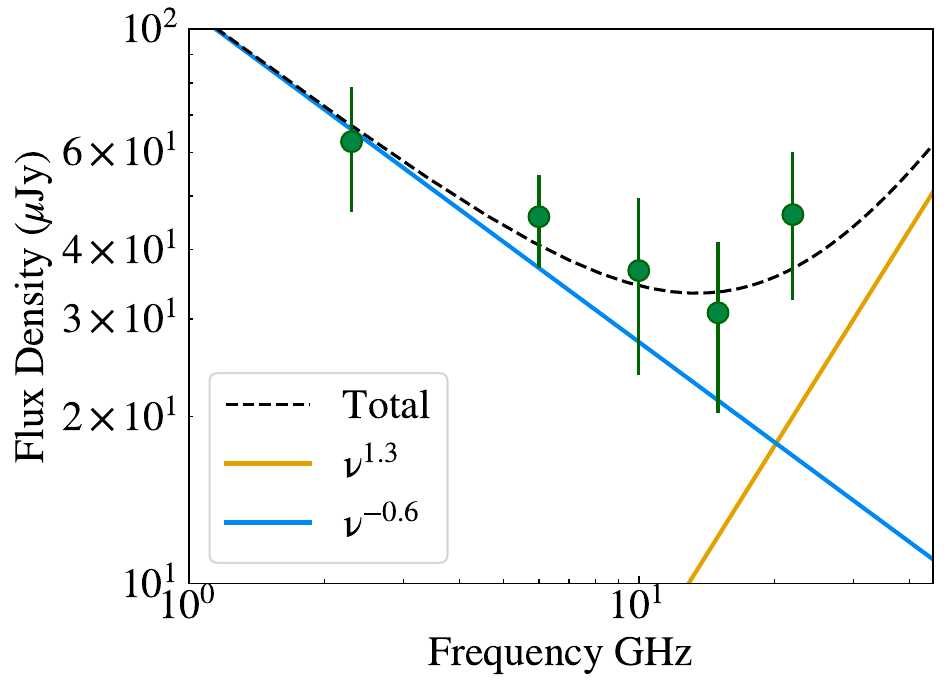}
    %\caption*{Two Powerlaw Fit}
%\end{minipage}
\caption{Two-powerlaw fit to the KISS15s radio SED. We note the inversion at $\sim$15 GHz, which clearly cannot be fit by a single powerlaw. A powerlaw with a negative exponent (corresponding to the blue line) is expected of synchrotron emission from SNe IIn in the optically thin regime, while a positive exponent may be indicative of synchrotron emission from an optically thick region. With no constraint on the peak of either component of the SED, a full FFA/SSA fit was not possible.}
\label{fig:k15s_r_fits}
\end{figure}

The spectral energy distribution of KISS15s (with flux densities measured as described in Section \ref{subsec:R_an}) is displayed in Figure \ref{fig:k15s_r_fits}. The spectral rise due to synchrotron emission is expected to follow a powerlaw, 

\begin{equation}
\mathrm{S(\nu)\propto\nu^{-\alpha}}
\label{eq:powerlaw}
\end{equation}

Where S is flux density, $\nu$ is frequency, and $\alpha$ is the spectral index. The spectral index for synchrotron emission from SNe IIn in the optically thin regime is expected to be $\gtrsim0.5$ \citep[e.g.][]{Ellison1991,Chevalier2017}. We find an unusual SED from KISS15s: the flux densities from 3-15 GHz follow a power-law with a negative exponent (as expected), but the 20 GHz point rises again, creating a spectral inversion. We fit a two component model to the KISS15s radio SED with MCMCs, finding that the spectrum is best reproduced by two powerlaws with spectral indices of \eq{\alpha=0.6_{-0.23}^{+0.24}} and \eq{\alpha=-1.3_{-0.20}^{+0.37}}, as shown in Figure \ref{fig:k15s_r_fits}. We note that there are likely degeneracies in this fit. Further fitting in a FFA/SSA \citep{Chevalier2003} context is not possible given the limited data points. We simply conclude that there is evidence for two-component radio emission, but do not have the sensitivity to capture the details of the radio emission. The spectral inversion is further discussed in Section \ref{subsec:spec_inversion}.

\begin{deluxetable*}{lcccc}
%\caption{Optical and X-ray Mass-Loss Rates for Our Sample and Other SNe IIn.}
\label{tab:MLR_Opt_Xray}
\tablewidth{0pt}
\tablehead{
\colhead{SN} &
\colhead{Optical Epoch (d)} &
\colhead{Optical ML Rate (\eq{M_{\odot}yr^{-1}})} &
\colhead{X-ray Epoch (d)} &
\colhead{X-ray ML Rate ($\rm{M_{\odot}yr^{-1}}$)}
}
\startdata
2005ip   & $\sim$3770 & $(1.50\pm0.75)\times10^{-4}$ & 750  & $(1.5 \pm 0.5) \times 10^{-2}$ \\
2010jl   & $\sim$84   & $0.9 \pm 0.45$  & 3664 & $(3.5 \pm 1.2)\times10^{-3}$ \\
2013L    & $\sim$300  & $0.08 \pm 0.07$, $0.0041 \pm 0.0039$& 4312 & $(2.0\pm1.3)\times10^{-3}$ \\
2014ab   & $\sim$150  & $1.0 \pm 0.5$, $0.3 \pm 0.15$   & 4042 & $<2\times10^{-3}$ \\
2015da   & $\sim$100  & $0.5 \pm 0.25$  & 3581 & $<1.7\times10^{-3}$ \\
KISS15s  & $\sim$300  & $0.4 \pm 0.2$   & 3313 & $(4.0 \pm 1.6)\times 10^{-3}$ \\
\enddata
\caption{Optical and X-ray Mass-Loss Rates for Our Sample and Other SNe IIn. Data not from this work taken from \cite{Kokubo2019,Andrews2017,Taddia2020,Smith2024,Bilinski2020,Katsuda_2014,Zhang_10jl,Chandra_2025,Smith_2017}. Optical mass-loss rates are reported as averages of range estimates/values that agree, or multiple values for those cases (SNe 2013L and 2014ab) in which values differ by $\sim$ an order of magnitude (due to different wind speeds/conversion efficiency assumptions). Errors on optical mass-loss rates are estimated as 50$\%$ due to the uncertainty in conversion efficiency of bolometric luminosity to H$\alpha$ luminosity, and incorporate additional uncertainty from large range estimates.}
\end{deluxetable*}

\needspace{2\baselineskip}

\section{Discussion}\label{sec:Discussion}
Using our late-time X-ray and radio observations, we estimated the mass-loss rate of each of the four supernovae centuries before explosion as detailed in Section \ref{sec:Results}. We show these mass-loss rates along with the optical estimates from early times in Table \ref{tab:MLR_Opt_Xray}. Our late-time multiwavelength analysis suggests a consistent drop in mass-loss rates for all four SNe as we probe earlier in the progenitor stars' late-stage evolution. This is particularly interesting as the early optical studies of these SNe suggested that they all belong to the same subgroup of 2010jl-like SNe IIn, characterized by high mass-loss rates and luminous, enduring light curves \citep{Andrews2017,Bilinski2020,Smith2024,Kokubo2019}. 

SN 2013L exhibited early evidence of pre-existing or newly formed dust, anisotropic CSM, and a mass-loss rate as high as \eq{\sim0.15~M_{\sun}\,yr^{-1}} in the decades prior to explosion \citep{Andrews2017, Taddia2020}. Our X-ray observations suggest ongoing interaction with lower-density CSM formed from mass-loss at a rate of \eq{\rm{\dot{M}}\approx2.0\times10^{-3}~M_{\sun}\,yr^{-1}} at \eq{\approx382\, yrs} pre-supernova (for a $\sim$120 km/s CSM speed).

Mass-loss rates on the order of \eq{10^{-3}~M_{\sun}\,yr^{-1}} long before explosion are only achieved by the steady winds of the most luminous stars known, including the most extreme LBVs like $\eta$~Carinae, whereas most LBV winds have mass loss rates an order of magnitude less \citep{smith26}. Thus, even rates around \eq{10^{-3}~M_{\odot}\,yr^{-1}} require some sort of enhanced or episodic mass-loss mechanism. There is no proposed single-star mechanism to account for the strong episodic and eruptive mass loss of LBVs, but, as stated, several clues suggest that LBV eruptions most likely arise from binary mergers or rapid mass-transfer events \citep{Smith2015,smith18}.

Timescales for binary Roche-lobe overflow (RLOF) are expected to be on the order of \eq{\sim10^{4}\,yr}, but this system could be explained by a binary that underwent partially conservative RLOF or lost mass in later binary evolutionary stages, such as a common envelope phase or pre-merger inspiral mass-loss \citep{Ivanova2013, Smith2014, Pejcha2017, Igoshev2020}. 

\begin{figure}
    \centering
    \includegraphics[width=8.7cm, height=6.8cm]{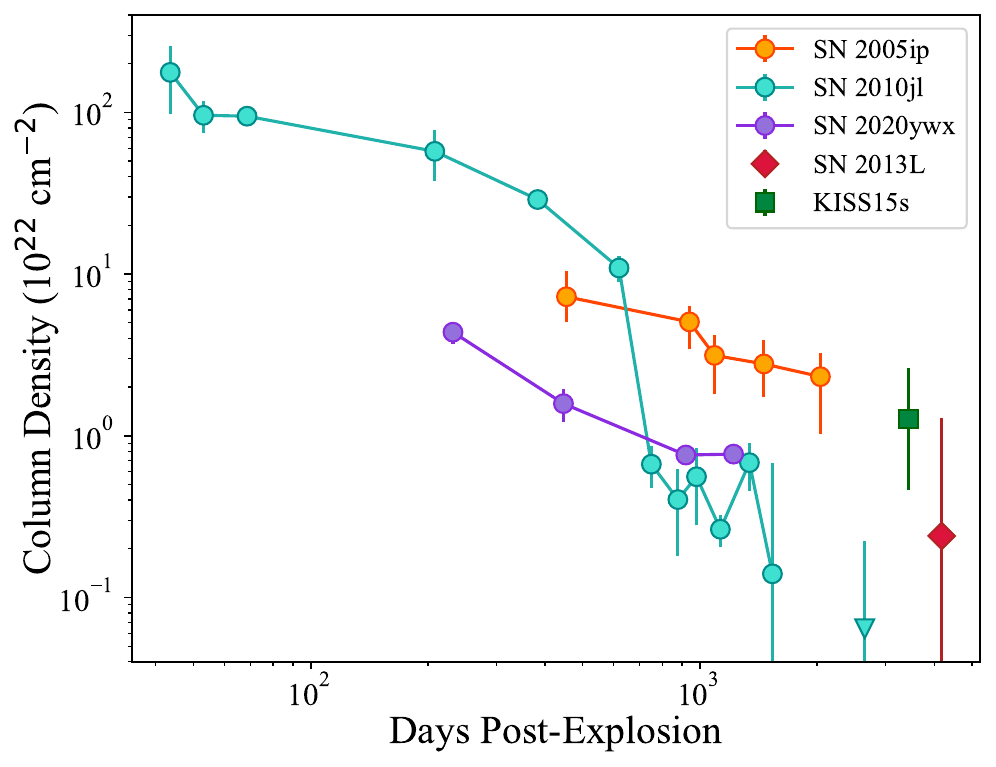}
    \caption{KISS15s and SN 2013L column density estimates compared with values obtained for SNe 2005ip, 2010jl, and 2020ywx at earlier epochs \citep[from][]{Katsuda_2014, Chandra2015, BaerWay2025}. Explosion dates for the SNe in this work are taken as optical discovery date, and number of days post-explosion are calculated from the observation time weighted average. KISS15s in particular shows a higher-than-expected column density at this late epoch, possibly due to an enduring cool dense shell.}
    \label{fig:Column_10jl15s}
\end{figure}

We found both the 0.2 - 10 keV luminosity and the column density of KISS15s to be unexpectedly high (see Figure \ref{fig:Column_10jl15s} for a comparison of the SN 2013L and KISS15s column densities with those of other bright, long-lasting SNe IIn). The spectral inversion in the radio emission from KISS15s further suggests we are potentially viewing a second shock as was seen in i.e. SN 1986J \citep{Bietenholz2002}. Our column density measurement at late times is higher than seen in SN 2010jl by more than an order of magnitude. Based on the X-ray luminosity, we determine a mass-loss rate of \eq{\dot M\approx4\times10^{-3}\,\rm{M_{\odot}~yr^{-1}}} centuries before the explosion. This is much lower than the rate found by \citealt{Kokubo2019} of $\sim 0.4\,\rm{M_{\odot}\,yr^{-1}}$ at early epochs, suggesting the mass-loss rate increased greatly in the final years of the progenitor's life. 

The continuously elevated mass-loss rate, combined with the asymmetries seen in the optical spectra taken by \citet{Kokubo2019}, suggests that KISS15s may have originated in a binary system where the stellar mass-loss ramped up by 1-2 orders of magnitude pre-supernova. 

For SNe 2014ab and 2015da, the upper limits are only somewhat constraining on the mass-loss rates given that the observations are at such late stages. We find, however, that the mass-loss rates are at least one to two orders of magnitude lower than those measured at earlier times --- and our upper limit on \eq{\dot M} is inconsistent with direct extrapolations from optical measurements in the case of SN~2015da --- suggesting again that the mass-loss intensified dramatically preceding the explosion. 

While it is possible that there are deviations from the equipartition we assume that could partially explain the discrepancies between early and late-time mass-loss rates, it seems that these objects can be explained through outbursts only in the last decades before the explosion. Wave-driving and gravity waves act on too short a timescale to explain the persisting mass-loss seen in previous observations of these systems \citep{Yoon_Cantiello,Quataert_2012}. The high mass-loss rates estimated at early times in all four SNe correspond to escalated mass-loss during the core carbon- and oxygen-burning phases, which may be associated with eruptive mass-loss episodes in the last few decades of a star's life. Pulsation-driven escalated mass-loss is proposed for single massive stars with initial masses $>$16 \Ms\ \citep{sengupta_2025}. In this evolution channel, shock build-up on the stellar surface drives dynamical mass ejection in the last centuries to decades before explosion, which leads to the formation of a very dense CSM close to the star at the time of explosion. 

\begin{figure}
    \centering
    \includegraphics[width=8 cm, height = 7
    cm]{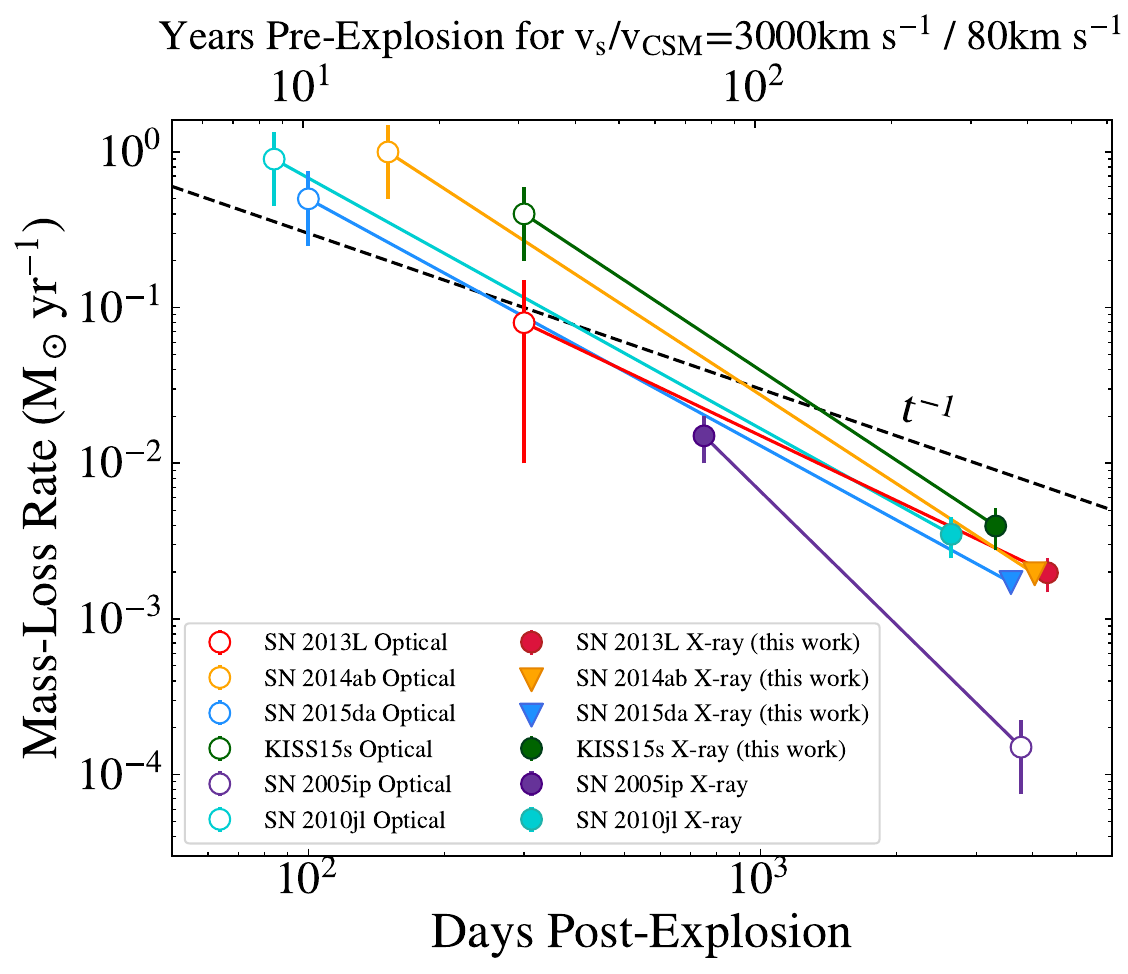}
    \caption{Our mass-loss results alongside earlier estimates for our sample and two other SNe IIn. X-ray results are shown with filled markers, while optical results are shown with open markers. We show a $t^{-1}$ curve to emphasize the sharp increase in mass-loss rate leading up to the supernova. Results for SN 2013L are from \citet{Taddia2013}, SN 2014ab from \citet{Bilinski2020}, SN 2015da from \citet{Smith2024} and KISS15s from \citet{Kokubo2019}. Results for SN 2010jl/SN 2005ip are from \citet{Zhang_10jl,Katsuda_2014,Smith_2017,Chandra_2025}. For this plot, we only include objects that have estimates at both $<$ 1000 and $>$ 3000 days post-explosion to understand long-terms trends. Errorbars on optical measurements are dominated by assumed 50$\%$ uncertainty in the conversion efficiency to H$\alpha$ luminosity.}
    \label{fig:Final_ML}
\end{figure}

In Figure \ref{fig:Final_ML}, we compare mass-loss rates and limits derived in this work with rates from early optical observations and those of SNe 2005ip and 2010jl, which shared similar early-time characteristics. The increase in mass-loss rate in all objects is steeper than $t^{-1}$, implying that most of the mass-loss happened in the final decades before explosion. The lack of exact constraints on the timeframe of elevated mass-loss precludes a precise estimate of the total CSM mass for each of these objects. We approximate, however, that the late-time results presented in this work suggest between \eq{5-20~M_{\odot}} of CSM in these objects. This would mean that all of these objects would have to come from massive stars \eq{>20-30~M_{\odot}} \citep{Smith2017}.

\subsection{Dust Formation}
Another intriguing aspect of these four supernovae is that they all showed an infrared excess. However, the lack of contemporaneous IR and radio/X-ray observations prevents a quantitative characterization of conversion efficiencies, clumping, or dust formation \citep{Chevalier2017}. The fact that these objects are quite distinct at radio and X-ray wavelengths suggests that IR excess at late times may be ubiquitous among SNe IIn simply because the dense shell provides an environment for dust formation \citep{Pozzo2004}, as seems to be indicated by the blueshifted line-profile evolution in SN~2015da \citep{Smith2024}. The X-ray detection of SNe 2013L and KISS15s gives some evidence that the interaction in these objects is ongoing and the dust formation could be occurring in the dense shell, but the X-ray data alone cannot distinguish between dust forming in the ejecta and the CDS (the two locations where dust can form post-explosion) \citep{Sarangi_2022}.

%\begin{figure}[H]
    %\centering
    %\includegraphics[width=.9\linewidth]%{15s_column_density.jpg}
    %\caption{Column density evolution of 2010jl compared to measured column density of KISS15s. Figure adapted with permission from \citet{Chandra2015}.}
   % \label{fig:k15s_nH}
%\end{figure}

Unlike SN~2010jl, SN~2015da and KISS15s showed no clear signature of IR excess in the first year post-explosion \citep{Smith2024, Kokubo2019}, which would originate from CSM dust formed before the explosion. High mass-loss rates facilitate dust formation in the pre-explosion wind. However, if the high mass-loss rates in the progenitors of these SNe were limited to only the last few decades prior to explosion, the dust would not have a chance to travel far from the star by the time of explosion. This dust might be evaporated by the SN flash \citep{fransson_2014,kochanek_2019,dwek_2021}. In KISS15s, the late-time radio detections suggest ongoing mass-loss for centuries pre-explosion, making the lack of early IR excess more difficult to explain, as presumably some of the CSM would be beyond the evaporation radius. It is plausible that this indicates that the CSM is aspherical, and that we are not seeing the illuminated infrared echo due to viewing angle effects. Monitoring across wavelengths at higher cadences is needed to explore this possibility in future objects. In any case, in SNe with a dense CSM, new dust forms in the interaction region between the forward and reverse shocks \citep{Smith2017, Sarangi_2022, shahbandeh_2024}. In SN~2015da and KISS15s, a clear blueshift in the H$\alpha$ emission lines is seen after one year post-explosion \citep{Smith2024, Kokubo2019}, indicating new dust formation in a dense shell behind the forward shock or in the ejecta. As Figure \ref{fig:Column_10jl15s} suggests, the forward shock in KISS15s is propagating through a relatively dense CSM even after 3000 days post-explosion. Even though the exact column density of the CSM behind the shock cannot be computed due to a lack of X-ray data at the earlier epochs, an estimate based on mass-loss rates suggests column densities larger than 10$^{24}$ cm$^{-2}$. A larger column density aids the formation of dust in the cool dense shell, by shielding X-rays and UV from the shock front \citep{sarangi_2018a}.

\subsection{Radio Spectral Inversion of KISS15s}\label{subsec:spec_inversion}
A radio spectral inversion such as that seen in KISS15s (Figure \ref{fig:k15s_r_fits}) implies a complex CSM geometry, most likely indicating multiple synchrotron emitting regions with different optical depths. Inverted radio spectra have been seen twice before in SNe IIn, in SN 1986J \citep{Bietenholz2002} and SN 2001em \citep{Chandra2020}. In the case of SN 1986J, \citet{Bietenholz2002} suggests the potential formation of a pulsar wind nebula (as the synchrotron optical depth from the interaction itself, $\tau_{\nu}$, drops below 1) moving at slow speeds \eq{\sim500~km\,s^{-1}}. VLBI monitoring of SN 1986J has confirmed that the central emission component has become dominant, but has not definitively answered whether we are seeing a pulsar wind nebula \citep{Bietenholz_2017}. However, \cite{Bietenholz_2017} determined that the inversion was not due to an absorption process alone. For SN 2001em \citep{Chandra2020}, a similar inversion was also interpreted to come from a central component, which may be a remnant of a dense mass ejection from common envelope evolution \citep{Chevalier2012}. Given the lack of data points available for the spectral inversion in KISS15s, it is not possible to firmly constrain what the inversion is caused by. We note that the X-ray measurements are not able to constrain the asymmetry expected in a common envelope scenario. However, we suggest that either the common envelope scenario or the pulsar wind scenario is plausible.

\section{Conclusion}\label{sec:Conclusion}

We present the results of a late-time multi-wavelength  (X-ray and radio) observational campaign of the four luminous and enduring SNe IIn SN 2013L, SN 2014ab, SN 2015da, and KISS15s. For all objects, we find evidence for an increasing mass-loss rate in the years leading up to explosion based on our derived rates and limits. For SN 2013L and KISS15s, we find evidence for continually elevated mass loss $>$ 300 years pre-explosion, suggesting a mass-loss mechanism that acted for many centuries and ramped up close to the explosion. Repeated LBV eruptions are plausible, likely due to binary interaction. For SNe 2014ab and 2015da, we find evidence for a more significant decline in mass-loss, suggesting that the elevated mass-loss may have only acted for a limited period. These results suggest that an increasing mass-loss rate in the years preceding the explosion seems to be a common feature of the SNe IIn subclass. Our work has increased the baseline of SNe IIn to late times. As shown by this study, late-time radio and X-ray observations are key to disentangling the variety within CSM-interacting supernovae, and further observations of similar samples to these will provide a more robust understanding of Type IIn SNe as well as the evolution and nature of their progenitors.

\section{Acknowledgments}
We thank the anonymous referee for thoughtful comments that improved the manuscript. 

EH is currently supported by the National Science Foundation Graduate Research Fellowship Program under Grant Number DGE-2545911. Any opinions, findings, and conclusions or recommendations expressed in this material are those of the authors and do not necessarily reflect the views of the National Science Foundation. 
RBW is supported by the National Science Foundation Graduate Research Fellowship Program under Grant number 2234693 and acknowledges support from the Virginia Space Grant Consortium.
PC acknowledges the support of this work provided by the National Aeronautics and Space Administration through Chandra Award Number GO4-25044X, along with GO3-24056X. 
KM is supported by Japan Society for the Promotion of Science (JSPS) KAKENHI grants JP24KK0070 and JP24H01810.

This research has made use of data obtained from the Chandra Data Archive provided by the Chandra X-ray Center (CXC). The National Radio Astronomy Observatory and Green Bank Observatory are facilities of the U.S. National Science Foundation operated under cooperative agreement by Associated Universities, Inc. We thank the staff of the VLA and the GMRT, who made these observations possible. GMRT is run by the National Centre for Radio Astrophysics of the Tata Institute of Fundamental Research.

This work made use of the following software packages: \texttt{HEASoft} \citep{HEASARC2014}, \texttt{CIAO} \citep{CIAO2006}, \texttt{CASA} \citep{McMullin2007}, \texttt{CARTA} \citep{CARTA2021}, \texttt{PIMMS} \citep{Mukai1993}, \texttt{matplotlib} \citep{Hunter:2007}, \texttt{numpy} \citep{numpy}, \texttt{pandas} \citep{mckinney-proc-scipy-2010,pandas_18675244}, \texttt{python} \citep{python}, \texttt{corner.py} \citep{corner-Foreman-Mackey-2016,corner.py_14209694}, \texttt{Cython} \citep{cython:2011}, \texttt{emcee} \citep{emcee-Foreman-Mackey-2013,emcee_10996751}, and \texttt{h5py} \citep{collette_python_hdf5_2014,h5py_7560547}.

This research has made use of the Astrophysics Data System, funded by NASA under Cooperative Agreement 80NSSC21M00561.

Software citation information aggregated using \texttt{\href{https://www.tomwagg.com/software-citation-station/}{The Software Citation Station}} \citep{software-citation-station-paper,software-citation-station-zenodo}.

% \input{appendix}


\bibliography{ref}{}
\bibliographystyle{aasjournal}



\end{document}